\documentclass[traditabstract]{aa}
\usepackage{placeins}
\usepackage{txfonts}
\usepackage{graphicx}
\usepackage{natbib}
\usepackage{aas_macros}
\usepackage{morefloats}
\usepackage{hyperref}
\hypersetup{
    colorlinks,
    citecolor=blue,
    filecolor=blue,
    linkcolor=blue,
    urlcolor=blue,
    menucolor=black}
\bibpunct{(}{)}{;}{a}{}{,}
\begin{document}

\newcommand{\blos}{B_{\rm los}}
\newcommand{\bthr}{B_{\rm thr}}
\newcommand{\bloc}{\left\langle{}B\right\rangle_{\rm loc}}
\newcommand{\cclv}{1-C\left(\mu=0.2\right)/C\left(\mu=1.0\right)}
\newcommand{\mumax}{\mu_{\rm max}}
\newcommand{\cmax}{C_{\rm max}}

\title{Intensity contrast of solar network and faculae II.  Implications for solar irradiance modelling}
\author{K.~L.~Yeo\inst{\ref{inst1}}\and N.~A.~Krivova\inst{\ref{inst1}}}
\institute{
Max-Planck-Institut f\"{u}r Sonnensystemforschung, Justus-von-Liebig-Weg 3, 37077 G\"{o}ttingen, Germany \\
\email{yeo@mps.mpg.de}
\label{inst1}
}
\date{Received <date> / Accepted <date>}
\abstract{We aim to gain insight into the effect of network and faculae on solar irradiance from their apparent intensity. Taking full-disc observations from the Solar Dynamics Observatory, we examined the intensity contrast of network and faculae in the continuum and core of the Fe I 6173 {\AA} line and 1700 {\AA}, including the variation with magnetic flux density, distance from disc centre, nearby magnetic fields, and time. The brightness of network and faculae is believed to be suppressed by nearby magnetic fields from its effect on convection. The difference in intensity contrast between the quiet-Sun network and active region faculae, noted by various studies, arises because active regions are more magnetically crowded and is not due to any fundamental physical differences between network and faculae. These results highlight that solar irradiance models need to include the effect of nearby magnetic fields on network and faculae brightness. We found evidence that suggests that departures from local thermal equilibrium (LTE) might have limited effect on intensity contrast. This could explain why solar irradiance models that are based on the intensity contrast of solar surface magnetic features calculated assuming LTE reproduce the observed spectral variability even where the LTE assumption breaks down. Certain models of solar irradiance employ chromospheric indices as direct indications of the effect of network and faculae on solar irradiance. Based on past studies of the Ca II K line and on the intensity contrast measurements derived here, we show that the fluctuations in chromospheric emission from network and faculae are a reasonable estimate of the emission fluctuations in the middle photosphere, but not of those in the lower photosphere. The data set, which extends from 2010 to 2018, indicates that intensity contrast was stable to about 3\% in this period.}
\titlerunning{Intensity contrast of solar network and faculae II}
\authorrunning{Yeo et al.}
\maketitle

\section{Introduction}
\label{introduction}

The variation in the solar radiation that enters the Earth's atmosphere is a key consideration in the climate change debate \citep{gray10}. For this reason, climate simulations require historical solar irradiance, defined as the Earthward solar radiative flux at one AU, as input. This is provided by models of solar irradiance variability because measurements only date back to 1978, and even then have gaps and gross instrumental issues \citep{ermolli13,kopp14}. The variation in solar irradiance at solar rotation to cycle timescales is believed to be dominantly driven by photospheric magnetism \citep{domingo09,yeo17b}. Specifically, it is thought to be the outcome of the changing prevalence and distribution of dark sunspots {and} bright network and faculae on the solar surface. Models based on this assumption have been successful in replicating most of the features in measured solar irradiance \citep{solanki13}. For example, the SATIRE-S model \citep{wenzler06,ball12,yeo14b} reproduces over 90\% of the observed variability in the wavelength-integrated or total solar irradiance since 1978. Of course, a firm understanding of the radiant properties of sunspots, network, and faculae is pertinent to the proper modelling of solar irradiance variability. While sunspot radiance is well constrained \citep{maltby86,collados94,unruh99}, the same cannot be said of network and faculae, mainly because of the challenge in observing such small-scale structures.

The radiant properties of network and faculae are often probed by taking simultaneous solar magnetograms and intensity images, and studying the variation in network and facular intensity contrast with distance from disc centre and magnetogram signal \citep{topka92,topka97,lawrence93,ortiz02,yeo13}. As magnetic flux tubes tend towards a surface normal orientation as a result of magnetic buoyancy, the viewing geometry is effectively a function of distance from disc centre. Above a certain minimum, magnetic flux tubes exhibit similar kilogauss magnetic flux densities regardless of size \citep{solanki93a,solanki99}. This means that the magnetogram signal at network and faculae is indicative of the magnetic filling factor. Moreover, the magnetic flux tube size rises on average with magnetogram signal \citep{ortiz02,yeo13}. To study the intensity contrast of network and faculae as a function of distance from disc centre and magnetogram signal is to study how their radiance varies with viewing geometry, magnetic filling factor, and flux tube size.

The \cite{topka92,topka97} and \cite{lawrence93} investigations were based on scans of various disc positions made with the Swedish Vacuum Solar Telescope \citep[SST,][]{scharmer03}. The scans varied in spatial resolution (0.3 to 0.6 arcsec) and wavelength, making it difficult to constrain the disc centre-to-limb variation (CLV) in intensity contrast. In addition, taken from the ground, SST observations suffer from atmospheric seeing.  \cite{ortiz02} used full-disc observations from the Michelson Doppler Imager on board the Solar and Heliospheric Observatory \citep[SoHO/MDI,][]{scherrer95}. These observations are not only seeing-free because they are taken from space, but cover the entire solar disc at a uniform if relatively poor spatial resolution (4 arcsec). This allowed the authors to derive an empirical relationship describing the variation in network and facular intensity contrast with distance from disc centre and magnetogram signal. In the various studies, the authors examined the intensity contrast in the nearby continuum of photospheric lines. The spectral line intensity contrast is also of interest, not least because studies have {found evidence} that both continuum and spectral line changes might be relevant to solar irradiance variability \citep{mitchell91,unruh99,preminger02}. However, early investigations into the line core intensity contrast of network and faculae were limited by insufficient data, such that they were unable to constrain either the CLV or the magnetogram signal dependence with confidence \citep{frazier71,walton87,lawrence91,title92}. Taking observations from the Helioseismic and Magnetic Imager on board the Solar Dynamics Observatory \citep[SDO/HMI,][]{scherrer12}, \cite{yeo13}, abbreviated here as Yea13, examined the intensity contrast of network and faculae in both the continuum and the core of the Fe I 6173 \AA{} line. HMI returns seeing-free full-disc observations of the magnetic flux density and the continuum and line core intensity at a uniform, intermediate spatial resolution (0.9 arcsec), allowing Yea13 to circumvent the limitations of the earlier studies. Using a simple model based on the observed CLV and magnetograms signal dependence of the continuum and line core intensity contrast, the authors confirmed that solar irradiance variability is indeed the sum manifest of continuum and spectral line changes.

Comparing network and facular intensity contrast by looking at quiet and active regions separately, \cite{lawrence93} and \cite{kobel11} noted that above a certain minimum magnetogram signal level, the quiet-Sun network is brighter than active region faculae at a similar magnetogram signal level. \cite{ortiz02} and Yea13 found that the intensity contrast per unit magnetic flux density decreases with increasing magnetic flux density. Taking into consideration that in their analyses, the points with lower magnetic flux densities correspond mainly to network and higher levels to faculae, both studies concluded that network is, on a per unit magnetic flux density basis, brighter than faculae. {The various investigations point to network having} a greater heating efficiency than faculae. This phenomenon is well supported by observations \citep{solanki84,solanki86,keller90} and magnetohydrodynamics (MHD) simulations \citep{vogler05,criscuoli13}. This has been attributed to the assumption that facular regions are more magnetically crowded; the more magnetically crowded a given area, the more convection is impeded and the less efficient the radiative heating of magnetic flux tubes by surrounding convection \citep{ishikawa07,kobel12,romano12,criscuoli13}. Significantly, by studying areas with varying degrees of magnetic activity, \cite{kobel12} found that the intensity contrast at a given magnetogram signal level and the Doppler velocity dispersion, which {is an indication} of convective strength, both decline gradually with the area-averaged magnetic flux density.

Here, we present a follow-up study to Yea13, who took HMI observations and examined the CLV and magnetogram signal dependence of network and facular intensity contrast in the continuum and core of the Fe I 6173 \AA{} line. Making use of similar HMI data and simultaneous observations from the Atmospheric Imaging Assembly \citep[AIA,][]{lemen12}, also on board SDO, we study the intensity contrast in the AIA 1700 \AA{} channel along that in the Fe I 6173 \AA{} line. In another advance, we examine not only the CLV and magnetogram signal dependence, but also the variation in intensity contrast with the average magnetic flux density over the neighbourhood of each network and facular image pixel, called here the local magnetic flux density, and time. The objectives of the study are listed below.
\begin{itemize}
        \item We aim to advance what Yea13 understood about network and facular radiance from the intensity contrast in the Fe I 6173 \AA{} line by a comparison to the contrast in the AIA 1700 \AA{} channel.
        \item \cite{kobel12} studied the disc centre. We extend this earlier study by examining the variation in intensity contrast with local magnetic flux density across the entire solar disc. {This} is to shed light on how the observation that network has a higher heating efficiency than faculae might be appropriately described in solar irradiance models.
        \item HMI and AIA have been in operation since 2010, almost the length of an entire solar cycle. We investigate if network and facular radiance might vary with cycle phase by looking at intensity contrast as a function of time. Clearly, this also has implications for the modelling of the effect of network and faculae on solar irradiance.
        \item Certain solar irradiance models adopt chromospheric indices as direct indications of the effect of network and faculae on solar irradiance \citep[e.g.][]{morrill11,thuillier12,coddington16,yeo17a}. Taking the intensity contrast measurements derived here and what is known about Ca II K radiance into consideration, we discuss the validity of this assumption.
\end{itemize}
In the following section, we describe the HMI and AIA observations we employed. Then, we detail how we isolate the network and faculae in the data set (Sect. \ref{analysis_fac}), and derive the corresponding intensity contrast (Sect. \ref{analysis_pix}) and local magnetic flux density (Sect. \ref{analysis_loc}). We present and discuss the results in Sect. \ref{results}, before giving a summary of the study in Sect. \ref{summary}.

\section{Data}
\label{data}

\subsection{HMI}
\label{hmi}

The HMI \citep{scherrer12} records full-disc polarised filtergrams at the Fe I 6173 \AA{} line at 1.875s intervals on two $4096\times4096$ pixel CCDs. The pixel scale and spatial resolution is 0.5 and 0.9 arcsec, respectively. A range of data products is generated from the filtergram sequence, including 45s and 720s cadence Dopplergrams, longitudinal magnetograms, and maps of the continuum intensity, and the depth and width of the Fe I 6173 \AA{} line. The line core intensity is given by the difference between the continuum intensity and line depth maps. Vector magnetograms are also generated at 720s cadence.

For this study, we made use of 45s longitudinal magnetograms, continuum, and line core intensity images. The continuum and core of the Fe I 6173 \AA{} line is formed in the lower photosphere \citep[around 20 km,][]{norton06} and over the middle photosphere \citep[the contribution function is centred on 250 km with a full width at half-maximum, FWHM, of 300 km,][]{fleck11}, respectively.

The longitudinal magnetogram gives the mean line-of-sight magnetic flux density over each resolution element, $\blos$. As magnetic flux tubes tend towards a surface normal orientation, we can approximate the pixel-averaged magnetic flux density, $B$ by the quotient $\blos/\mu$, where $\mu$ is the cosine of the heliocentric angle. The $1\sigma$ noise level of HMI 45s longitudinal magnetograms and 720s vector magnetograms is about 10 G \citep{liu12} and 100 G \citep{hoeksema14}. The elevated noise level of HMI vector magnetograms, arising from the uncertainty in measured Stokes Q and U, is the reason why we choose to estimate $B$ from $\blos$ over taking it directly from vector magnetograms.

\subsection{AIA}
\label{aia}

The AIA \citep{lemen12} comprises four $4096\times4096$ pixel CCD telescopes that image the Sun in ten wavelength channels. The pixel scale and spatial resolution are 0.6 and 1.5 arcsec. Seven channels are centred on extreme ultraviolet (EUV) lines that are formed in the transition region and corona. The other three channels are at nominally 1600 \AA{}, 1700 \AA{}, and 4500 \AA{}. Full-disc images are recorded in each EUV channel at 12s intervals, the 1600 \AA{} and 1700 \AA{} channels at 24s intervals, and the 4500 \AA{} channel hourly.

Here, we examine the intensity contrast of network and faculae in the 1700 \AA{} channel. The 1700 \AA{} channel captures continuum emission, mainly from neutral silicon bound-free transitions, that forms over the middle photosphere \citep[the response function is centred on 360 km with an FWHM of 325 km,][]{fossum05}.

We excluded the other AIA channels from this study for the following reasons. The EUV channels are formed {in the transition region and corona,} where magnetic structures are no longer co-spatial to their photospheric footpoints, and are therefore not suitable for such a study. The 1600 \AA{} channel captures continuum emission that forms in the upper photosphere and C IV emission that forms in the transition region \citep{handy99,lemen12}, which complicates {the interpretation of} the apparent intensities. AIA 4500 \AA{} images have {severe} artefacts that are caused by damage to a filter.\footnote{Reported by Dean Parnell on the official SDO mission weblog, \url{sdoisgo.blogspot.de}, in the entry dated 17 December 2013.}

\subsection{Data selection and reduction}
\label{reduction}

\begin{figure*}
\centering
\includegraphics[width=17cm]{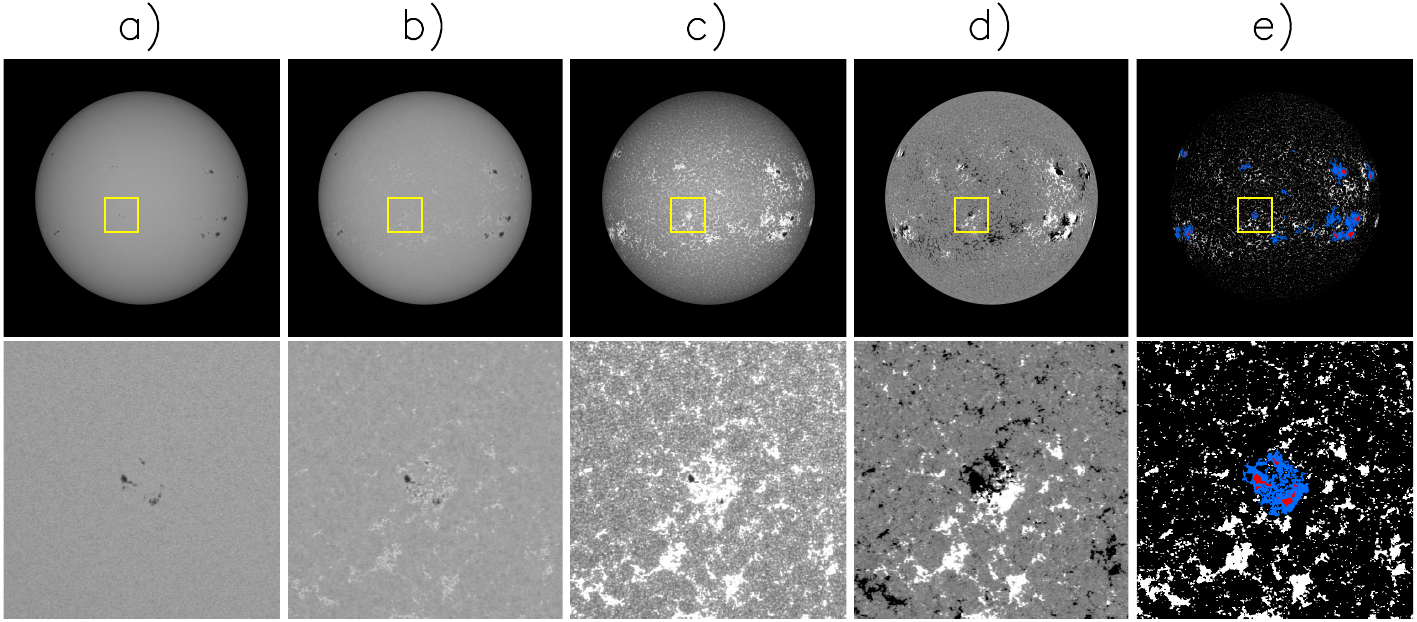}
\caption{Top: HMI and AIA observations from one of the data days, 12 June 2014. From left to right, the intensity image at a) the continuum and b) the core of the Fe I 6173 \AA{} line, and c) the AIA 1700 \AA{} channel, d) the longitudinal magnetogram, and e) the map of the image pixels identified as network and faculae (white). The image pixels identified as sunspots are marked in red. The magnetic activity contiguous to sunspots and pores, marked in blue, is excluded from the analysis (see Sect. \ref{analysis_fac}). The greyscale is saturated for the intensity images at null and 150\% of the quiet-Sun level at disc centre, and at $\pm50\ {\rm G}$ for the magnetogram. Bottom: Corresponding zoomed-in inset of the boxed $300\times300$ arcsec region.}
\label{detection}
\end{figure*}

HMI and AIA have been in continuous operation since April 2010. We made use of data from 100 days between June 2010 and October 2018, selecting from each calendar month the day with the highest sunspot number \citep{clette16}.

For each data day, we identified a 315s period over which HMI and AIA operated without interruption. We took the seven HMI 45s longitudinal magnetograms and continuum and line core intensity images from this interval, and the 11 AIA 1700 \AA{} images recorded over the same period. We rotated each magnetogram to the middle of the 315s period and took the average, likewise for the other data products. The result is a 315s longitudinal magnetogram and intensity image at each passband, which are co-temporal (Fig. \ref{detection}). This averaging suppresses the variability from noise, p-mode oscillations \citep{krijger01}, and acoustic waves excited in the convection zone \citep{wedemeyer04}.

We co-registered the HMI and AIA observations using the aia\_prep.pro routine (SolarSoft IDL). This resamples the HMI data to the AIA image pixel scale. It is known that the co-registration by this routine is only accurate to about 1 arcsec because of short-term variability in the pointing of AIA and HMI, possibly from thermal flexing, which is not captured in the data header \citep[see][and references therein]{briceorange14}. We refined the alignment between the HMI and AIA data by comparing the absolute magnetogram signal to the 1700 \AA{} intensity at disc centre, taking care to exclude sunspots, and finding the shift to the magnetogram that maximises the cross-correlation between the two. The required shift varied between 0.3 and 1.2 arcsec over the data days.

\section{Analysis}
\label{analysis}

\subsection{Identification of network and faculae}
\label{analysis_fac}

Network and faculae were identified following Yea13.
First, we determined the noise level of the HMI 315s longitudinal magnetograms, $\sigma_{\blos}$ as a function of position on the solar disc by the exact procedure described in Yea13. (We cannot employ the noise level estimate from this earlier study here as it was determined for HMI observations at the original spatial resolution.) The noise level increases radially, from an average of 4.5 G around disc centre ($0.9<\mu\leq1.0$) to 6.4 G near the limb ($0.1<\mu\leq0.2$).

For each data day, we took the points on the solar disc where $\blos>3\sigma_{\blos}$ in the longitudinal magnetogram to harbour magnetic activity. The weakest magnetic flux density detectable, $\bthr$ is given by $3\sigma_{\blos}/\mu$. This rises from 14 G around disc centre to 127 G near the limb from the combination of the disc centre-to-limb increase in the noise level and foreshortening. The continuum intensity image is normalised by the quiet-Sun level and flat-fielded (detailed next in Sect. \ref{analysis_pix}). The active pixels where the normalised continuum intensity is below 0.89 are identified as sunspots. The remaining active pixels are taken as network and faculae, with the following exclusions (see detailed discussion in Yea13):
\begin{itemize}
        \item Yea13 noted that the magnetic canopy of sunspots extend beyond their boundary in the continuum image. As in that study, we excluded the active pixels contiguous to sunspots. Although some of these points might correspond to network and faculae, the apparent magnetic flux density is affected by the encroaching magnetic canopy and stray light originating from the sunspots. This measure also accounts for bright penumbra filaments that are not captured by the continuum intensity criterion.
        \item Stand-alone active pixels are excluded to minimise the incorrect inclusion of magnetogram noise as network and faculae.
        \item Active pixels where $B>800\ {\rm G}$ are excluded because they mainly correspond to bright features associated with sunspots near the limb than network and faculae. {See the detailed discussion in Yea13.}
        \item Close to the limb, the $B=\blos/\mu$ approximation breaks down from the combination of magnetogram noise and foreshortening. For this reason, we ignored the active pixels where $\mu\leq0.1$. The effect on the study is very light because only 1\% of the solar disc, by area, lies outside the $\mu=0.1$ loci.
\end{itemize}
It is worth emphasising here that it is not necessary in such an investigation to include the network and faculae as completely as possible but rather to minimise false positives. While these measures inevitably result in some network and faculae being excluded, they minimise false positives while still retaining most of the network and faculae (Fig. \ref{detection}). A total of $5.0\times10^7$ image pixels are identified as corresponding to network and faculae. They span $\bthr<B\leq800\ {\rm G}$ and $0.1<\mu\leq1.0$, the widest range of magnetic flux densities and disc positions we can reasonably examine with HMI observations.

\subsection{Intensity contrast}
\label{analysis_pix}

Following Yea13, the intensity contrast at position ($x,y$) in a given image $C\left(x,y\right)$ is defined as
\begin{equation}
C\left(x,y\right)=\frac{I\left(x,y\right)}{I_{\rm qsn}\left(x,y\right)}-1,
\label{intensitycontrast}
\end{equation}
where $I$ and $I_{\rm qsn}$ denote the apparent intensity and the quiet-Sun level, respectively.

Taking each intensity image, we masked the points where $B>10\ {\rm G}$ in the corresponding longitudinal magnetogram, leaving the quiet Sun. The quiet-Sun intensity as a function of distance from disc centre, that is, the limb-darkening profile, is given by the fifth-order polynomial in $\mu$ fit to this quiet-Sun image \citep[following][]{neckel94}. We normalised the quiet-Sun image by the limb-darkening profile. Image distortions from inhomogeneities in instrumental properties across the image plane manifest in the normalised quiet-Sun image as deviations in the overall level from unity. The flat-field image, mapping these image distortions, is given by the polynomial surface fit to the normalised quiet-Sun image (after first smoothing it with a $401\times401$ pixel boxcar filter to filter out small-scale variability). For the 1700 \AA{} images, where the image distortions are most severe, the root mean square (RMS) difference between the flat-field image and unity ranged from $3.2\%$ to $4.6\%$ over the data days. We corrected the original intensity image for image distortions by {dividing it by the flat-field image} before calculating the intensity contrast (Equation \ref{intensitycontrast}). We determined the limb-darkening profile and flat-field image of each intensity image separately on the observation that HMI and AIA instrument throughput and the image distortions appear to vary with time.

The line core intensity is modulated not just by fluctuations in line strength and shape, but also by continuum excess that fills in the spectral line. To aid the interpretation of line core intensity contrasts, we factored out continuum changes by dividing each line core intensity image by the corresponding normalised continuum intensity image before we calculated the intensity contrast. (With this normalisation, the line core intensity image indicates the level if the continuum is fixed at the quiet-Sun level.) We refer to Yea13, who examined the effect of this step on apparent line core intensity contrasts in detail.

The data reduction (Sect. \ref{reduction}) and the process by which we isolated network and faculae (Sect. \ref{analysis_fac}), and determined their intensity contrast, just described, closely follows Yea13. {The only significant departure from Yea13 is that we resampled the HMI observations to the image pixel scale of AIA in order to extend the study to include the AIA 1700 \AA{} channel (Sect. \ref{reduction}).} As we demonstrate in Sect. \ref{results_pix_a1_er}, this had negligible effect on the apparent intensity contrast in the HMI passbands.

\subsection{Local magnetic flux density}
\label{analysis_loc}

One of the aims of this study is to examine the effect of how magnetically crowded a given region of the solar surface is on the radiant properties of the enclosed network and faculae. To this end, we studied the variation in intensity contrast with what we call the local magnetic flux density, $\bloc$, which is defined as the mean absolute magnetic flux density in the $9\times9$ pixel ($5.4\times5.4$ arcsec) window centred on each network and facular image pixel. To minimise any bias from magnetogram noise, we smoothed each longitudinal magnetogram with a Gaussian kernel with an FWHM of 3 pixels before we calculated this quantity.

We adopted a window size of $9\times9$ pixel. Tests showed that the window size has no qualitative effect on the results, to be presented in Sect. \ref{results_pix_b}. The effect of magnetic concentrations on one another evidently diminishes with separation such that only the magnetic fields closest to a given point is relevant. Therefore broadening the window beyond a certain minimum does not add more of the relevant neighbouring magnetic activity, but merely scatters the resulting $\bloc$.

\section{Results}
\label{results}

\subsection{Variation with disc position and magnetic flux density}
\label{results_pix_a1}

\begin{figure*}
\centering
\includegraphics[width=16cm]{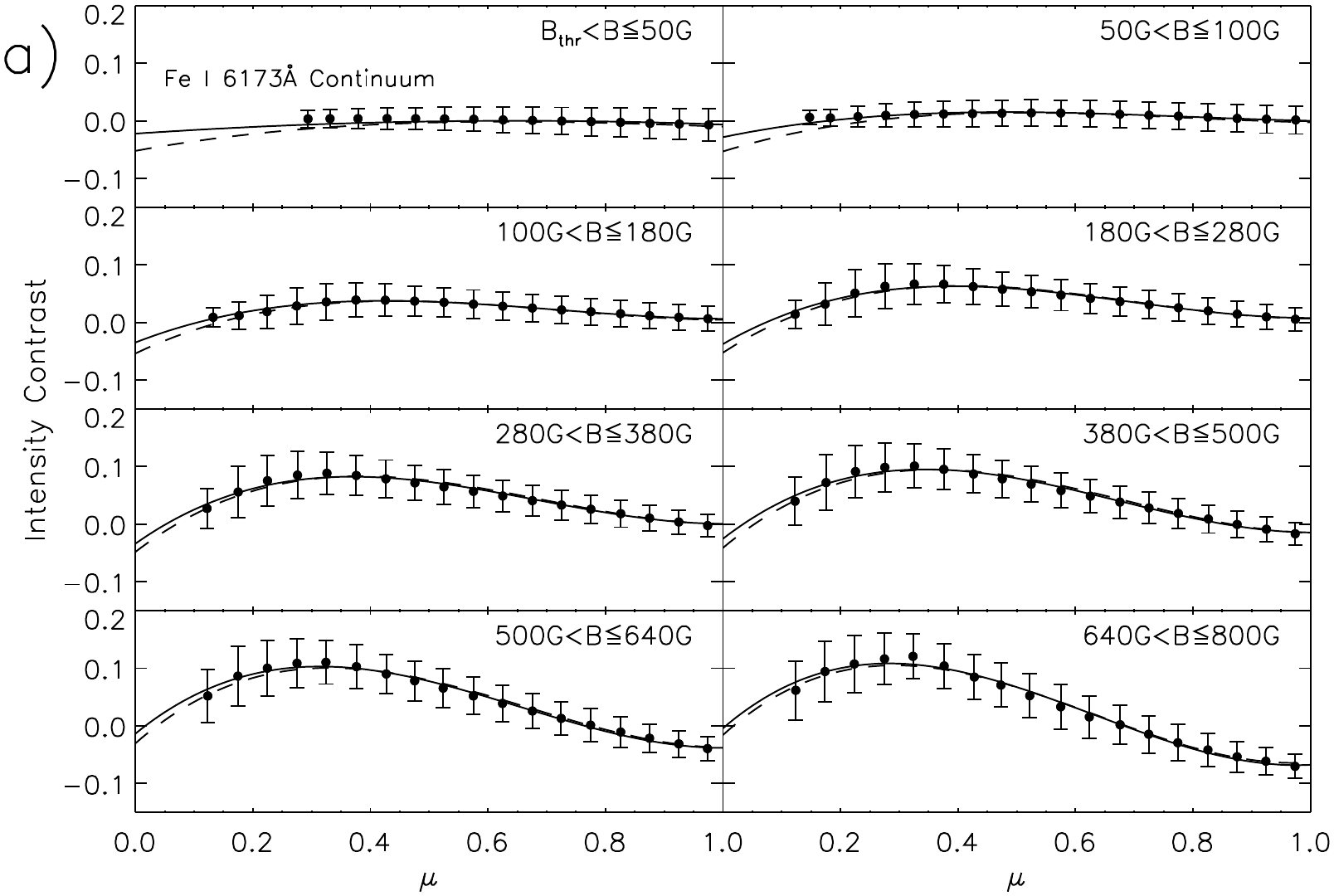}
\includegraphics[width=16cm]{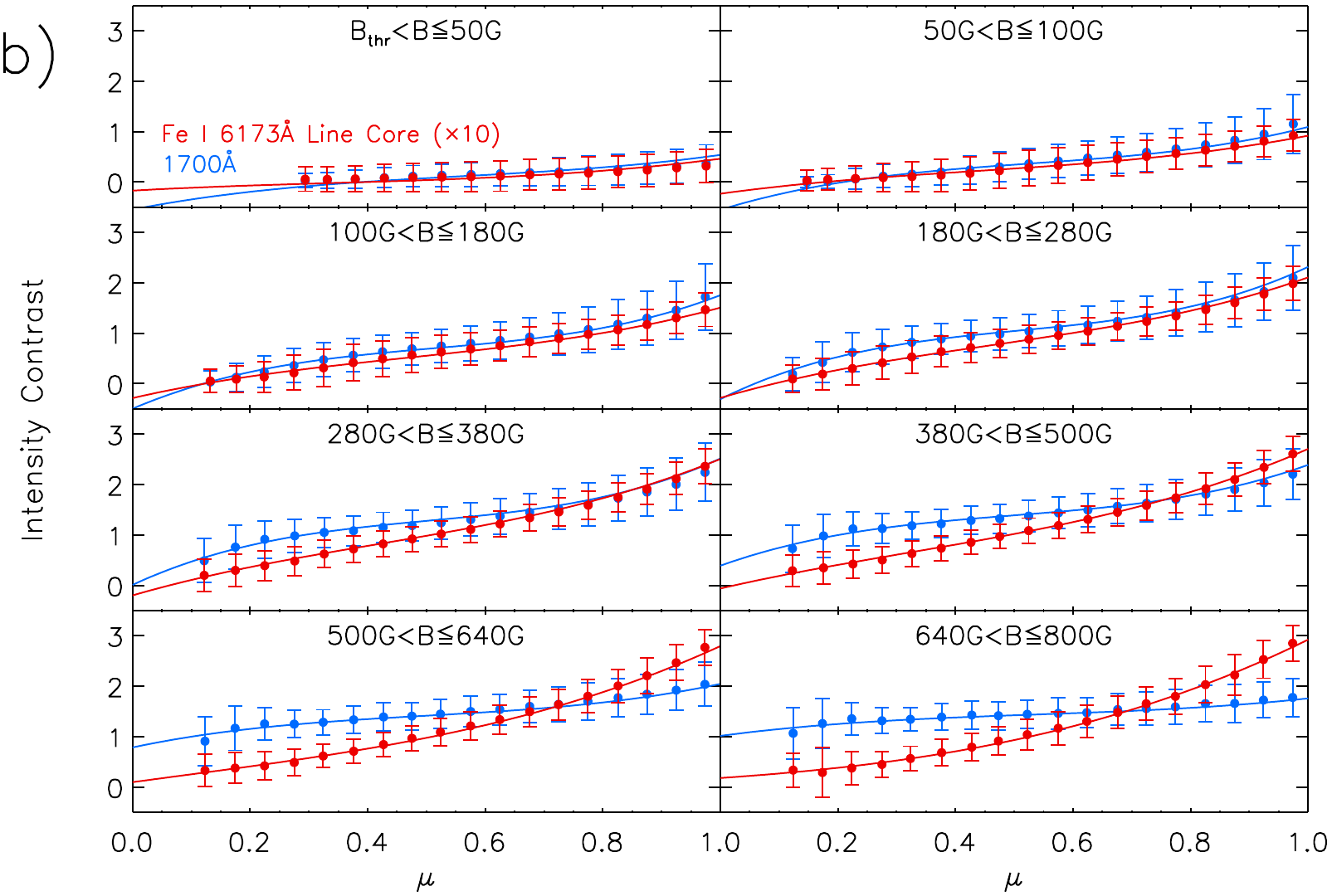}
\caption{a) Intensity contrast at the continuum of the Fe I 6173 \AA{} line in the annotated intervals of $B$. We binned the intensity contrast measurements within each $B$ interval by $\mu$ in bins of 0.05. The filled circles and error bars correspond to the mean and standard deviation within each $\mu$ bin. {The standard error of the mean is given by the quotient of the standard deviation and the square root of the number of measurements in the bin. Each bin contains $10^5$ measurements, such that the standard error is too small to be represented in this plot.} The solid curves follow the bivariate polynomial in the $\mu$ and $B$ fit to the measured intensity contrast (Equation \ref{surfacefiteqn} and Table \ref{surfacefit}). The dashed curves represent the result from the similar analysis in Yea13 at the original spatial resolution of HMI, which are largely hidden because they agree so closely. b) The same, except for the intensity contrast at the core of the Fe I 6173 \AA{} line (red) and 1700 \AA{} (blue). The Fe I 6173 \AA{} line core values are scaled by a factor of 10 to facilitate the comparison.}
\label{contrastbinbybmu}
\end{figure*}

\begin{figure*}
\centering
\includegraphics[width=16cm]{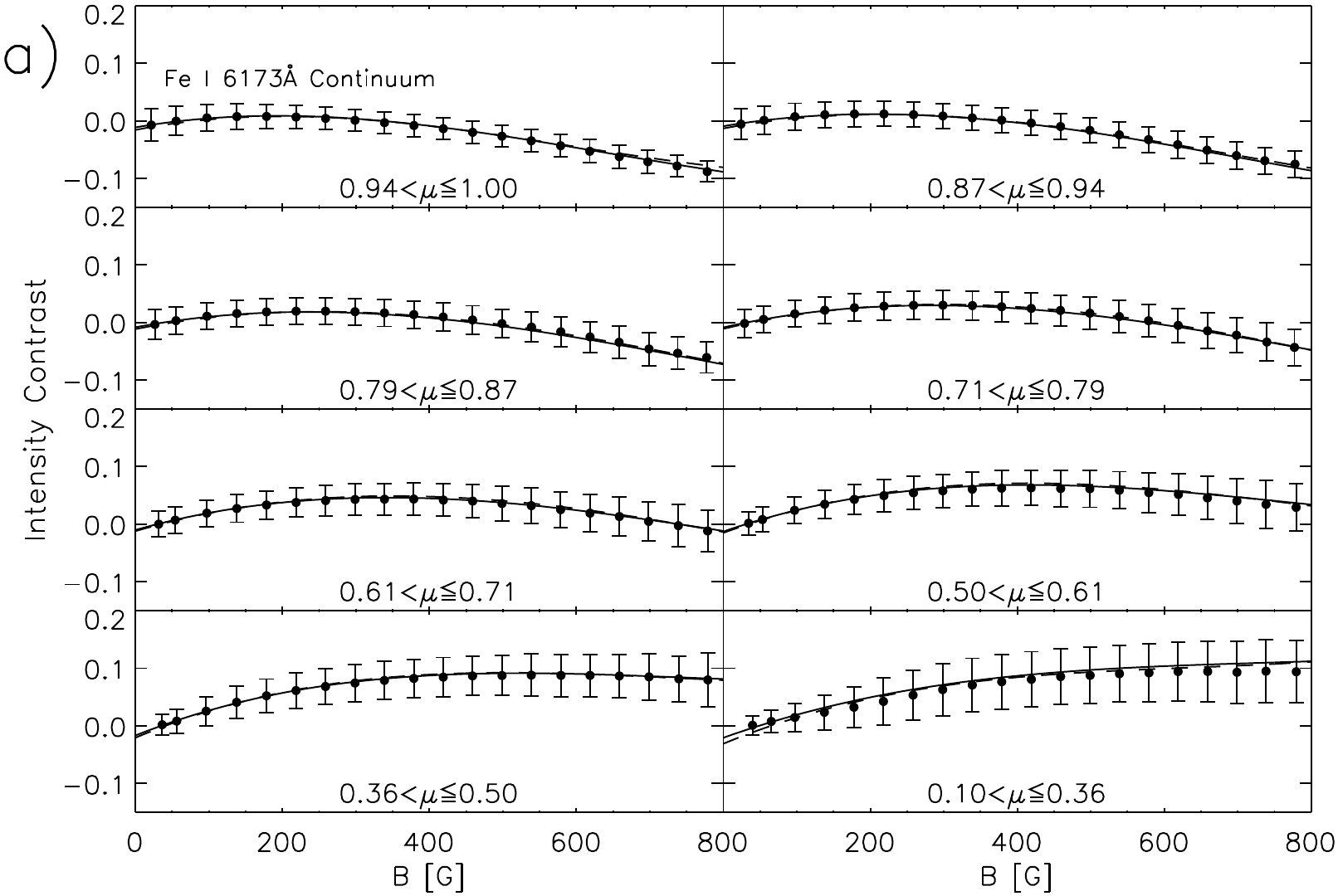}
\includegraphics[width=16cm]{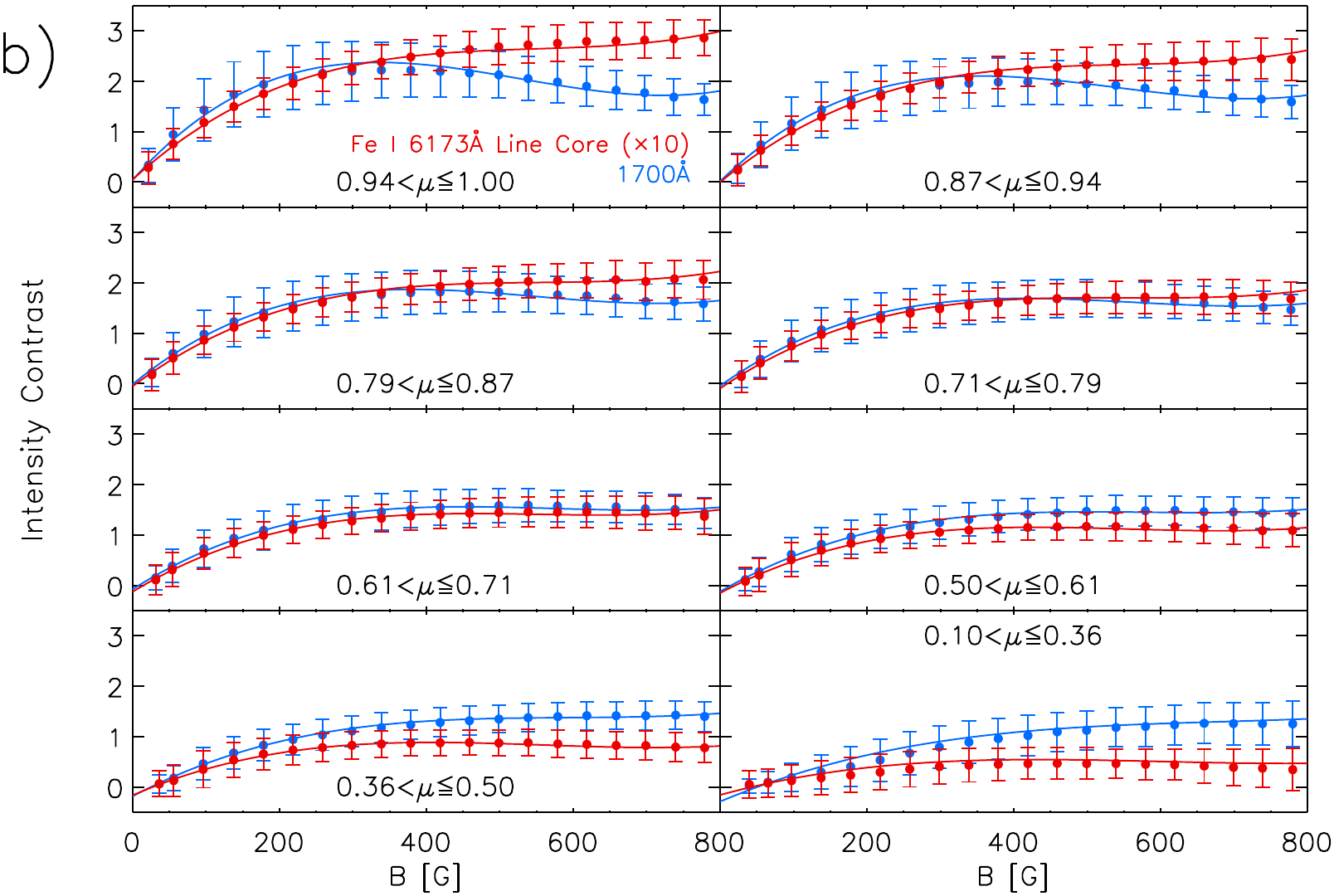}
\caption{a) Intensity contrast at the continuum of the Fe I 6173 \AA{} line as a function of $B$ from about disc centre ($0.94<\mu\leq1.00$) to near the limb ($0.10<\mu\leq0.36$). At each interval of $\mu$, we binned the intensity contrast measurements by $B$ in 40 G bins. The mean and standard deviation within each $B$ bin is represented by the filled circles and error bars. {As in Fig. \ref{contrastbinbybmu}, the standard error of the mean is negligible because of the size of our data set.} The solid curves represent the bivariate polynomial in $\mu$ and $B$ fit to measured intensity contrast (Equation \ref{surfacefiteqn} and Table \ref{surfacefit}) and the dashed curves show the same from Yea13, which is almost indistinguishable because of the close agreement. b) The corresponding plots for the core of the Fe I 6173 \AA{} line (red) and 1700 \AA{} (blue). Again following Fig. \ref{contrastbinbybmu}, we scaled the Fe I 6173 \AA{} line core values by a factor of 10.}
\label{contrastbinbymu}
\end{figure*}

\begin{figure}
\resizebox{\hsize}{!}{\includegraphics{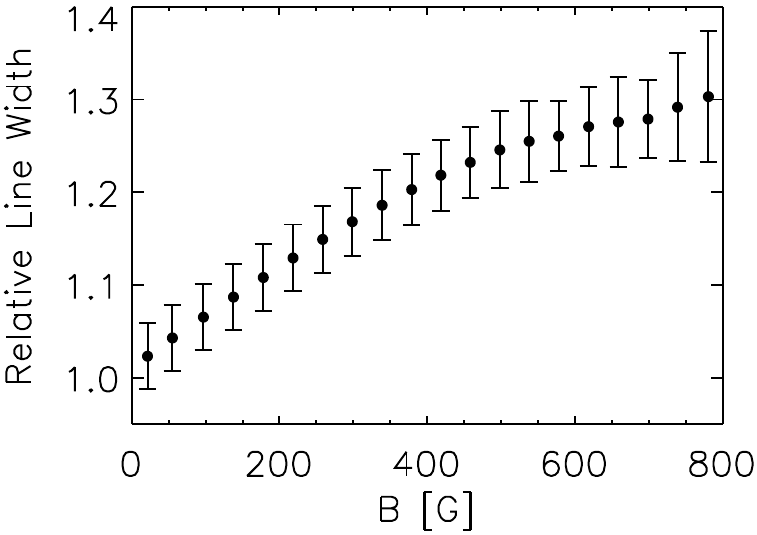}}
\caption{As a function of $B$, the width of the Fe I 6173 \AA{} line, relative to the quiet-Sun level, about disc centre ($0.94<\mu\leq1.00$). The relative line width of network and faculae, which is essentially their contrast in the line width map, is derived in an analogous manner to the intensity contrast (Sect. \ref{analysis_pix}). The filled circles and error bars have the same meaning as in Fig. \ref{contrastbinbymu}.}
\label{contrastlinewidth}
\end{figure}

\begin{figure}
\resizebox{\hsize}{!}{\includegraphics{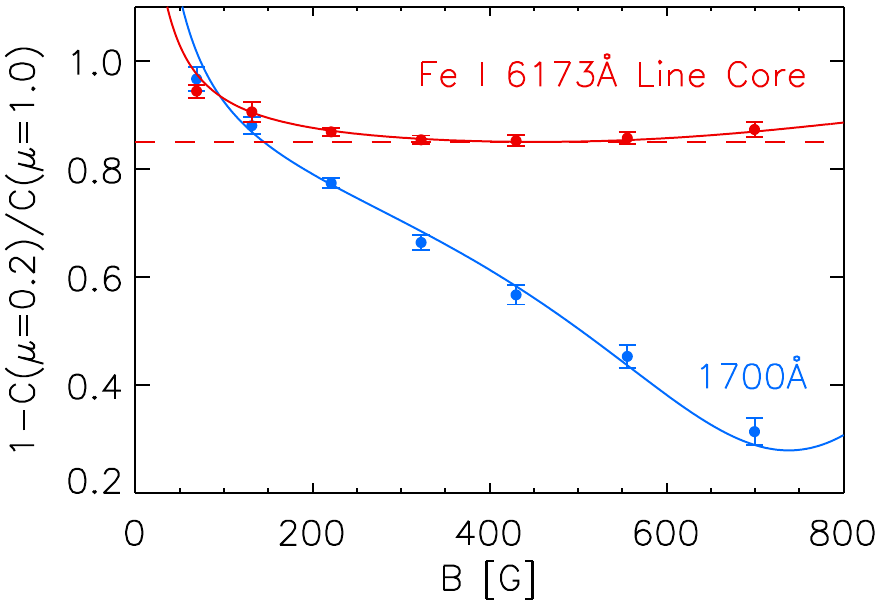}}
\caption{Proportional drop in intensity contrast between disc centre and $\mu=0.2$, $\cclv$ at the Fe I 6173 \AA{} line core (red) and 1700 \AA{} (blue). The plot points denote the estimate from the intensity contrast CLV profiles in Fig. \ref{contrastbinbybmu}b. The curves follow the solution from the empirical intensity contrast models (Equation \ref{surfacefiteqn} and Table \ref{surfacefit}). The dashed line marks the minimum of the Fe I 6173 \AA{} line core curve.}
\label{contrastclv}
\end{figure}

\begin{table*}
\caption{Coefficients of the bivariate polynomial in the $\mu$ and $B$ (Equation \ref{surfacefiteqn}) fit to the measured intensity contrast.}
\label{surfacefit}
\centering
\begin{tabular}{lcccccccc}
\hline\hline
 & $a_{00}$ & $a_{01}$ & $a_{02}$ & $a_{03}$ & $a_{10}$ & $a_{11}$ & $a_{12}$ & $a_{13}$ \\
Passband & $[\times10^{-2}]$ & $[\times10^{-2}]$ & $[\times10^{-2}]$ & $[\times10^{-2}]$ & $[\times10^{-3}]$ & $[\times10^{-3}]$ & $[\times10^{-3}]$ & $[\times10^{-3}]$ \\
\hline
Fe I 6173 \AA{} continuum & -1.68 & -5.09 & 16.5  & -11.1 & -0.217 & 4.30 & -7.80 & 3.95 \\
Fe I 6173 \AA{} line core & -1.08 & -2.26 & 0.844 & 3.22  & -0.236 & 3.21 & -3.69 & 2.11 \\
1700 \AA{}                & -51.8 & 105   & -77.1 & 28.9  & -1.31  & 45.0 & -78.9 & 52.9 \\
\hline
 & $a_{20}$ & $a_{21}$ & $a_{22}$ & $a_{23}$ & $a_{30}$ & $a_{31}$ & $a_{32}$ & $a_{33}$ \\
Passband & $[\times10^{-6}]$ & $[\times10^{-6}]$ & $[\times10^{-6}]$ & $[\times10^{-6}]$ & $[\times10^{-9}]$ & $[\times10^{-9}]$ & $[\times10^{-9}]$ & $[\times10^{-9}]$ \\
\hline
Fe I 6173 \AA{} continuum & 0.648 & -7.40 & 12.4 & -6.41 & -0.451 & 4.51 & -8.19 & 4.53 \\
Fe I 6173 \AA{} line core & 0.859 & -8.29 & 10.4 & -5.46 & -0.661 & 5.60 & -7.16 & 3.71 \\
1700 \AA{}                & 12.7  & -127  & 221  & -146  & -11.0  & 91.0 & -158  & 103  \\
\hline
\end{tabular}
\end{table*}

First, we examine the variation in intensity contrast with distance from disc centre and magnetic flux density, represented by $\mu$ and $B$, respectively. The intensity contrast measurements cover $0.1<\mu\leq1.0$ and $\bthr<B\leq800\ {\rm G}$. To elucidate the CLV, we grouped the measurements by $B$ into eight intervals and charted the trend with $\mu$ within each interval (Fig. \ref{contrastbinbybmu}). As for the $B$ dependence, we divided the solar disc into eight concentric annuli of equal area and examined the variation in intensity contrast with $B$ within each annulus (Fig. \ref{contrastbinbymu}).

\subsubsection{Empirical relationship}
\label{results_pix_a1_er}

Following \cite{ortiz02} and Yea13, we also encapsulated the CLV and $B$ dependence of intensity contrast at each passband into an empirical relationship given by the bivariate polynomial in $\mu$ and $B$ fit to the intensity contrast measurements. We binned the intensity contrast measurements by $\mu$ and $\ln{}B$ in intervals of 0.02 and 0.05, respectively. We then fit a cubic polynomial in $\mu$ and $B$ to the bin-averaged intensity contrast. In other words, we described the intensity contrast by a relationship of the form
\begin{eqnarray}
C\left(\mu,B\right)=
\left[\begin{array}{c}
B^0 \\ B^1 \\ B^2 \\ B^3
\end{array}\right]^T
\left[\begin{array}{cccc}
a_{00} & a_{01} & a_{02} & a_{03} \\ a_{10} & a_{11} & a_{12} & a_{13} \\ a_{20} & a_{21} & a_{22} & a_{23} \\ a_{30} & a_{31} & a_{32} & a_{33}
\end{array}\right]
\left[\begin{array}{c}
\mu^0 \\ \mu^1 \\ \mu^2 \\ \mu^3
\end{array}\right],
\label{surfacefiteqn}
\end{eqnarray}
where $a_{ij}$ denotes the coefficient of the $B^{i}\mu^{j}$ term, tabulated in Table \ref{surfacefit}. This captures the variation in intensity contrast with $\mu$ and $B$, as demonstrated by the alignment to the measured intensity contrast (Figs. \ref{contrastbinbybmu} and \ref{contrastbinbymu}).

In order to include the AIA 1700 \AA{} channel in this study, we resampled the HMI observations to the image pixel scale of AIA (Sect. \ref{reduction}). Figs. \ref{contrastbinbybmu}a and \ref{contrastbinbymu}a show that the fit to the measured intensity contrast at the continuum of the Fe I 6173 \AA{} line (solid curves) is almost exactly similar to that from the same analysis in Yea13 with HMI observations at the original spatial resolution (dashed curves). We noted the same at the Fe I 6173 \AA{} line core, which is not shown to avoid cluttering the plots. Resampling the HMI observations to the AIA image pixel scale had no appreciable effect on the apparent intensity contrasts.

\subsubsection{Fe I 6173 \AA{} line continuum}
\label{results_pix_a1_int}

The continuum of the Fe I 6173 \AA{} line is formed in the lower photosphere, where magnetic flux tubes are heated through the side walls of the opacity depression by radiation from the surrounding convection \citep{spruit76}. The intensity of network and faculae is modulated by the competing influence of this lateral heating and the magnetic suppression of convection within magnetic concentrations \citep{spruit81,grossmanndoerth94}.

With increasing distance from disc centre to limb, the intensity contrast increases to a maximum before declining towards the limb (Fig. \ref{contrastbinbybmu}a). The initial climb corresponds to the hot side wall of the opacity depression rotating into view and the cool centre becoming increasingly occulted. The side wall is eventually blocked as well, giving the drop near the limb. The average magnetic flux tube size increases with $B$ (\citealt{ortiz02}; Yea13), with the consequence that lateral heating becomes less and less efficient. This contributed to the observation here that the slope of the contrast-versus-$B$ profile diminishes with $B$ (Fig. \ref{contrastbinbymu}a).

\subsubsection{Fe I 6173 \AA{} line core and 1700 \AA{}}
\label{results_pix_a1_lcraia1700}

The core of the Fe I 6173 \AA{} line and the AIA 1700 \AA{} channel are formed over the middle photosphere, where magnetic flux tubes are heated by radiation from deeper layers \citep{knoelker91}, and mechanical and Ohmic dissipations \citep{moll12}. Mechanical heating includes longitudinal tube waves (LTWs) that {are} excited in magnetic flux tubes by surrounding convection \citep{carlsson92,fawzy12}. The LTWs steepen as they propagate upwards, eventually forming shocks and dissipating their energy. This is one avenue through which the temperature structure and therefore the radiant properties of magnetic flux tubes can be influenced by the inhibition of convection in their surroundings by nearby magnetic fields (c.f. Sect. \ref{results_pix_b}).

The Fe I 6173 \AA{} line core intensity is also modulated by Zeeman line changes and continuum excess filling in the spectra line, although the latter is factored out here in the derivation of intensity contrast (Sect. \ref{analysis_pix}). The effect of Zeeman line changes on intensity contrast is indicated by the following: The HMI line width maps concurrent to the HMI observations examined here are available for 27 of the 100 data days. Taking these data, we found that the line width at network and faculae increases monotonically with $B$, depicted for disc centre ($0.94<\mu\leq1.00$) in Fig. \ref{contrastlinewidth}. Because the Zeeman effect does not change the equivalent width of a given spectral line (except if it is saturated, which is not the case here), an increase in line width has to be accompanied by a decrease in line depth. Therefore, the steady rise in line width with $B$ indicates a similar trend in line core intensity.

As opposed to what we observed at the continuum of the Fe I 6173 \AA{} line (Fig. \ref{contrastbinbybmu}a), at these two passbands, the intensity contrast declines steadily from disc centre to limb (Fig. \ref{contrastbinbybmu}b). We quantify the strength of this decline by the proportional drop between disc centre and $\mu=0.2$, $\cclv$ (Fig. \ref{contrastclv}). For each interval of $B$ in Fig. \ref{contrastbinbybmu}b, we determined $\cclv$ from the cubic polynomial fit to the CLV profile (except for the $\bthr<B\leq50\ {\rm G}$ interval, as no data are available around $\mu=0.2$ here). The uncertainty is propagated from the regression error. We also estimated this quantity from the empirical intensity contrast models (Equation \ref{surfacefiteqn} and Table \ref{surfacefit}).

At 1700 \AA{} (blue, Fig. \ref{contrastclv}), $\cclv$ declines steadily with $B,$ while at the Fe I 6173 \AA{} line core (red), this quantity falls less steeply with $B$ and only up to about 400 G before it rises instead. The following physical processes are at work:
\begin{enumerate}
        \item {Magnetic flux tubes tend towards a surface normal orientation. This means, going from disc centre to the limb, rays change from passing through single flux tubes to passing in and out of multiple flux tubes} \citep{bunte93,solanki98}. This leads to a drop in intensity contrast as radiation from flux tubes {becomes} absorbed into the interleaving non-magnetised atmosphere that is traversed by the rays. Of course, the greater the magnetic filling factor, the weaker this effect.
        \item The temperature gradient of the solar atmosphere is weaker within magnetic concentrations \citep[e.g.][]{vernazza81,fontenla91} such that the temperature contrast increases with height. As both {the Fe I 6173 \AA{} line and the AIA 1700 \AA{} channel} are optically thick, {going} from disc centre to limb, {we look shallower and shallower into the solar atmosphere}, resulting in an increase in {temperature and intensity contrast. Obviously, this effect becomes stronger with magnetic filling factor. This mechanism counters the decline in intensity contrast from disc centre to limb (Fig. \ref{contrastbinbybmu}b) such that it is less steep than it would be without this effect.}
        \item The Fe I 6173 \AA{} line weakens from disc centre to the limb \citep{norton06}, which also decreases the contribution by Zeeman line changes to the intensity contrast. As Zeeman line changes become stronger with $B$, so does the disc centre to limb decline in intensity contrast from this effect.
\end{enumerate}
As $B$ and therefore the magnetic filling factor rises, the first two mechanisms lead to a weaker {disc centre-to-limb decline in intensity contrast}, while the third mechanism, which is only applicable to the Fe I 6173 \AA{} line core, does the opposite. The action of the first two mechanisms at 1700 \AA{} produces the monotonic decline in $\cclv$ with $B$. The competition between the first two mechanisms and the third at the Fe I 6173 \AA{} line core {produces the reversal in trend in $\cclv$ with $B$ at around 400 G.}

At about the disc centre ($0.94<\mu\leq1.00$, Fig. \ref{contrastbinbymu}b), the intensity contrast rises with $B$ but at a diminishing rate. While the intensity contrast at the Fe I 6173 \AA{} line core (red) increases monotonically with $B$, at 1700 \AA{} (blue) it peaks at about 350 G before it starts to decline. We explain the diminishing slope and the divergent trend above 350 G next in Sect. \ref{results_pix_a2_pi}. From disc centre ($0.94<\mu\leq1.00$, Fig. \ref{contrastbinbymu}b) to the limb ($0.10<\mu\leq0.36$), the $B$ dependence of the intensity contrast changes. At the Fe I 6173 \AA{} line core (red), the intensity contrast changes from increasing monotonically with $B$ to peaking at intermediate levels of $B,$ while the converse is apparent at {1700 \AA{} (blue)}. The reason is the $B$ dependence of the CLV at the two passbands (Fig. \ref{contrastclv}), as discussed above.

\subsection{Heating efficiency of the magnetic flux}
\label{results_pix_a2_pi}

\begin{figure}
\resizebox{\hsize}{!}{\includegraphics{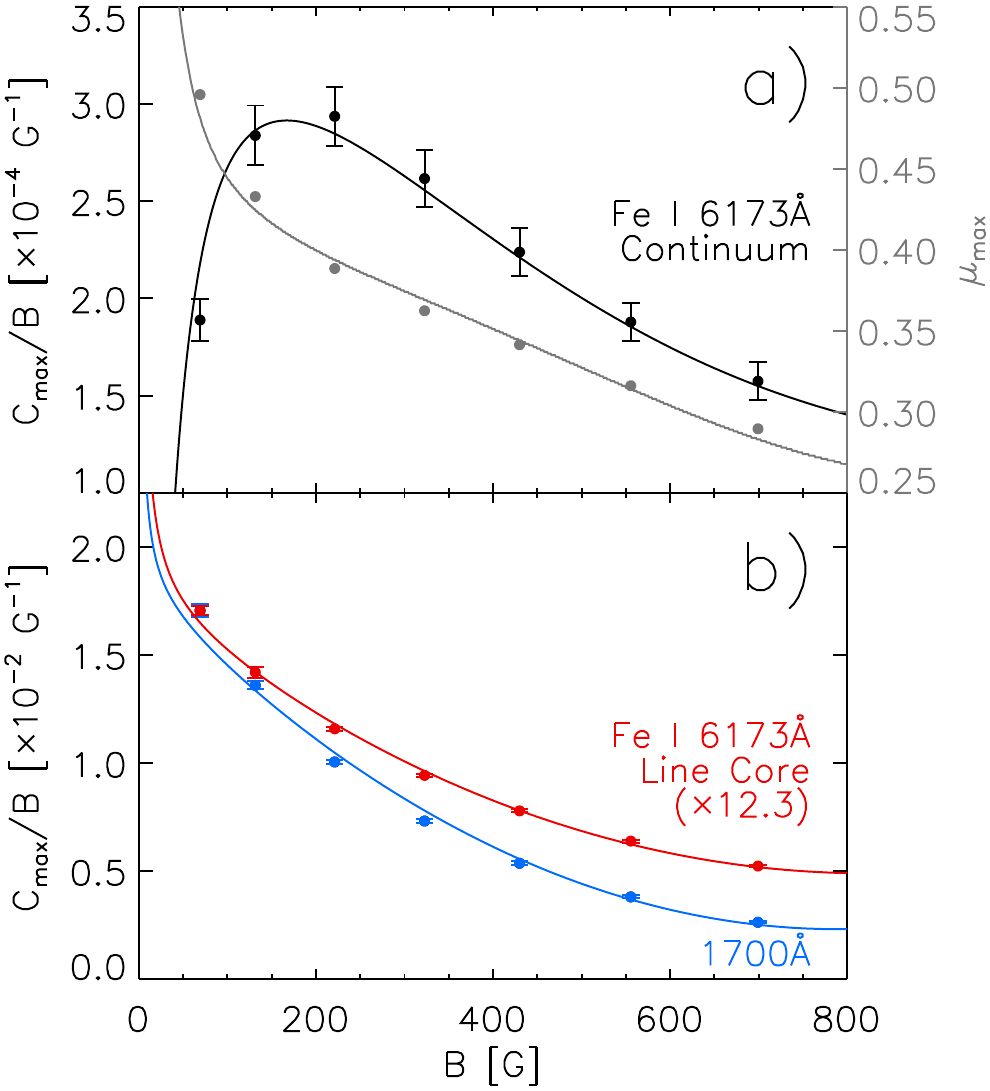}}
\caption{a) As a function of $B$, specific contrast, $\cmax/B$ (black, left-hand axis) and the position of $\cmax$, $\mumax$ (grey, right-hand axis) at the continuum of the Fe I 6173 \AA{} line. The plot points correspond to the values derived from the CLV profiles (Fig. \ref{contrastbinbybmu}a) and the curves to the estimate from the empirical intensity contrast models (Equation \ref{surfacefiteqn} and Table \ref{surfacefit}). b) Specific contrast at the Fe I 6173 \AA{} line core (red) and 1700 \AA{} (blue). To facilitate the comparison, we scaled the Fe I 6173 \AA{} line core values by a factor of 11.9 to match the 1700 \AA{} values at the first plot point. At these two passbands, $\mumax=1$ at all levels of $B$, hence the omission from the figure.}
\label{contrastspecific}
\end{figure}

We denote with $\mumax(B)$ and $\cmax(B)$  the position of and intensity contrast at the peak of the intensity contrast CLV profile at a given level of $B$. Following \cite{ortiz02} and Yea13, we examined the heating efficiency of magnetic flux as indicated by the intensity contrast per unit magnetic flux density at the CLV profile peak, $\cmax/B$, which we call the specific contrast.

At each passband and each interval of $B$ in Fig. \ref{contrastbinbybmu}, we estimated $\mumax$ and $\cmax$ from the cubic polynomial fit to the CLV profile. The specific contrast is then given by the quotient of $\cmax$ and the mean $B$ of the network and facular points in the $B$ interval. The corresponding uncertainty is propagated from the regression error and the {standard error of the mean $B$}. We excluded the $\bthr<B\leq50\ {\rm G}$ interval, where this analysis is adversely affected by the fact that the apparent intensity contrast and $B$ are weakest and therefore most uncertain here, and extend a limited range of $\mu$. We also determined $\mumax$ and $\cmax/B$ from the empirical intensity contrast models (Equation \ref{surfacefiteqn} and Table \ref{surfacefit}). {The results are depicted in Fig. \ref{contrastspecific}.}

At the continuum of the Fe I 6173 \AA{} line, the intensity contrast peaks closer and closer to the limb with increasing $B$ (grey, Fig. \ref{contrastspecific}a). This was similarly noted by \cite{ortiz02} and Yea13, who asserted, based on the hot wall model \citep{spruit76}, that this indicates the size of magnetic flux tubes on average increases with $B$. At the two other passbands, $\mumax=1.0$ at all levels of $B$ (Fig. \ref{contrastbinbybmu}b). The divergence stems from the difference in formation height and therefore the physical processes that drive the magnetic enhancement of intensity, as discussed in Sect. \ref{results_pix_a1}.

The specific contrast at the continuum of the Fe I 6173 \AA{} line (black, Fig. \ref{contrastspecific}a) rises with $B$, up to a maximum at about 150 G, before it starts to decline. At the Fe I 6173 \AA{} line core (red, Fig. \ref{contrastspecific}b) and 1700 \AA{} (blue), the specific contrast{, given here by the disc centre intensity contrast per unit $B$,} declines monotonically with $B$ and more rapidly so at the latter. {As noted in Sect. \ref{results_pix_a1_lcraia1700}, around disc centre ($0.94<\mu\leq1.00$, Fig. \ref{contrastbinbymu}b), the Fe I 6173 \AA{} line core (red) and the 1700 \AA{} intensity contrast (blue) therefore rises with $B$ at a diminishing rate and more rapidly so at 1700 \AA,{} such that the trend with $B$ reverses at around 350 G.}

As argued by Yea13, in non-sunspot magnetic features in full-disc observations, the points with lower magnetogram signals mainly correspond to the network, and the higher levels correspond to faculae. Therefore, the decline in specific contrast with $B$ indicates that the quiet-Sun network  for the same amount of magnetic flux is brighter and hotter than active region faculae. For a given magnetic flux tube, radiative heating through the side walls of the opacity depression and mechanical heating by LTWs are inhibited when convection in the greater area is suppressed by nearby magnetic fields. Because facular regions evidently contain more magnetic fields than the quiet Sun, this effect is greater on faculae than in the network. We examine this in greater detail in the following section (Sect. \ref{results_pix_b}). Another factor that could have contributed to the downward trend in specific contrast with $B$ at the continuum of the Fe I 6173 \AA{} line is the effect of the increase in average flux tube size with $B$ on the efficiency of the lateral radiative heating.

The divergent $B$ dependence at the continuum and core of the Fe I 6173 \AA{} line below 150 G was similarly noted by Yea13, who attributed it to the tendency of magnetic concentrations to aggregate in dark intergranular lanes on the intensity contrast in the lower photosphere, that is, the continuum \citep{title96,schnerr11}. The steeper decline with $B$ at 1700 \AA{} as compared to the Fe I 6173 \AA{} line core is at least partly due to the contribution by the Zeeman line changes to the intensity contrast at the latter, which increases with $B$ (Sect. \ref{results_pix_a1_lcraia1700}).

\subsection{Effect of the local magnetic flux}
\label{results_pix_b}

\begin{figure*}
\centering
\includegraphics[width=16cm]{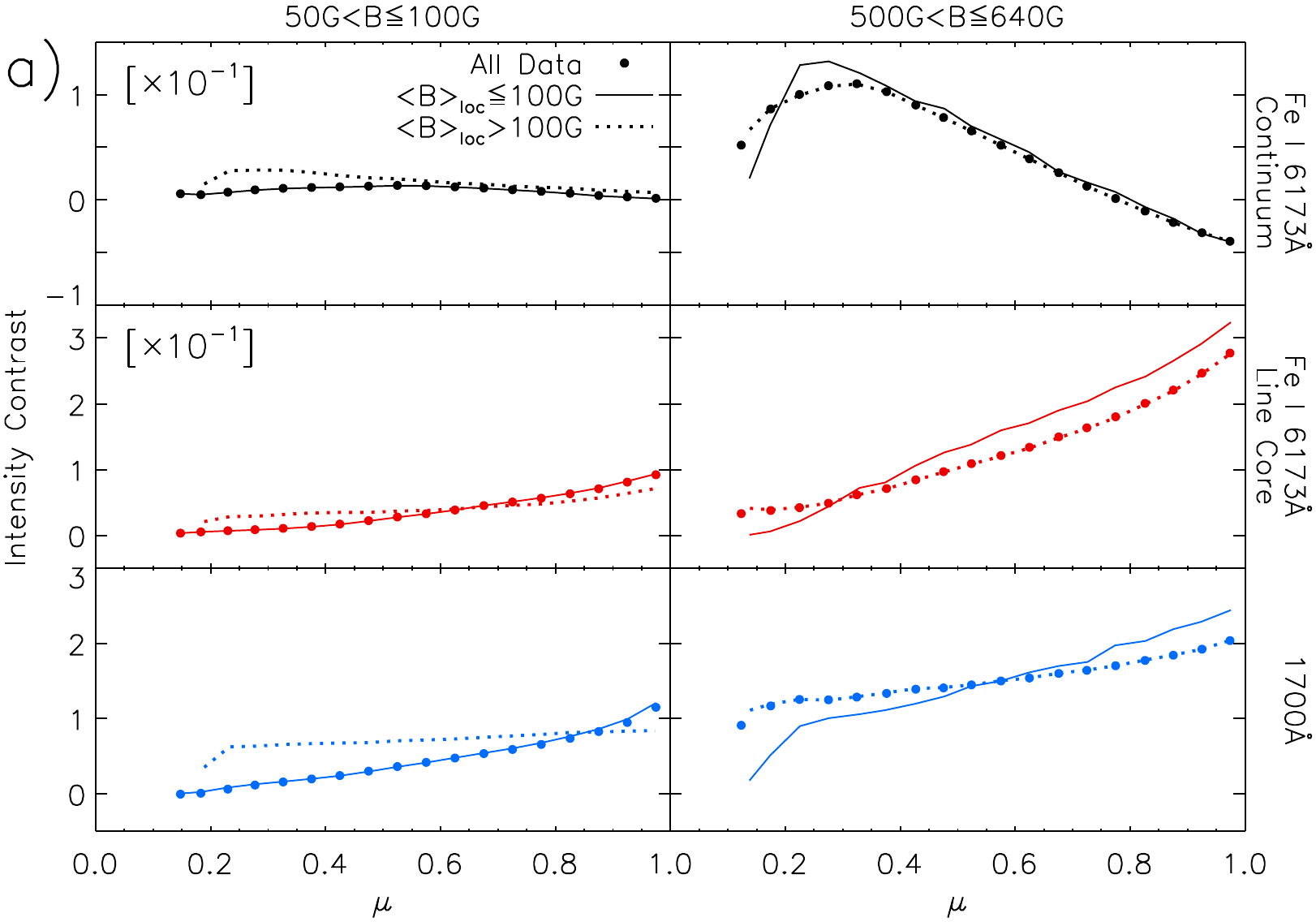}
\includegraphics[width=16cm]{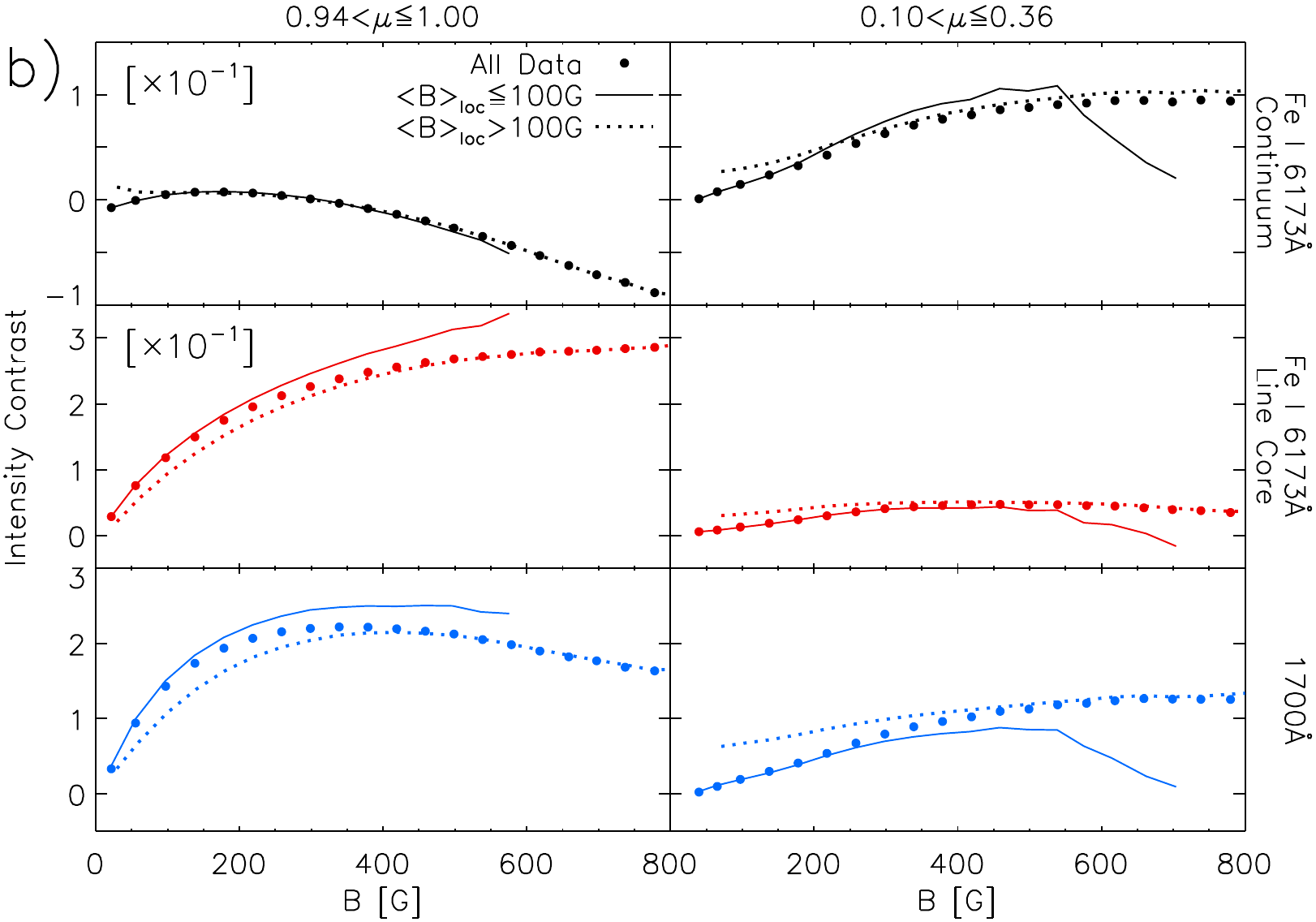}
\caption{a) CLV of the intensity contrast at low (left) and high $B$ (right), at $\bloc\leq100\ {\rm G}$ (solid curves) and $\bloc>100\ {\rm G}$ (dotted curves). From top to bottom, we show the values from the continuum (black) and core (red) of the Fe I 6173 \AA{} line and 1700 \AA{} (blue). b) Variation in intensity contrast with $B$ about disc centre (left) and near the limb (right), in the same $\bloc$ ranges. The intensity contrast profiles are derived in the same manner as the corresponding profiles, which were computed taking all available data into account (filled circles, taken from Figs. \ref{contrastbinbybmu} and \ref{contrastbinbymu}). {As in Figs. \ref{contrastbinbybmu} and \ref{contrastbinbymu}, the standard error is negligible because of the size of the data set.}}
\label{contrastnbmu}
\end{figure*}

\begin{figure}
\resizebox{\hsize}{!}{\includegraphics{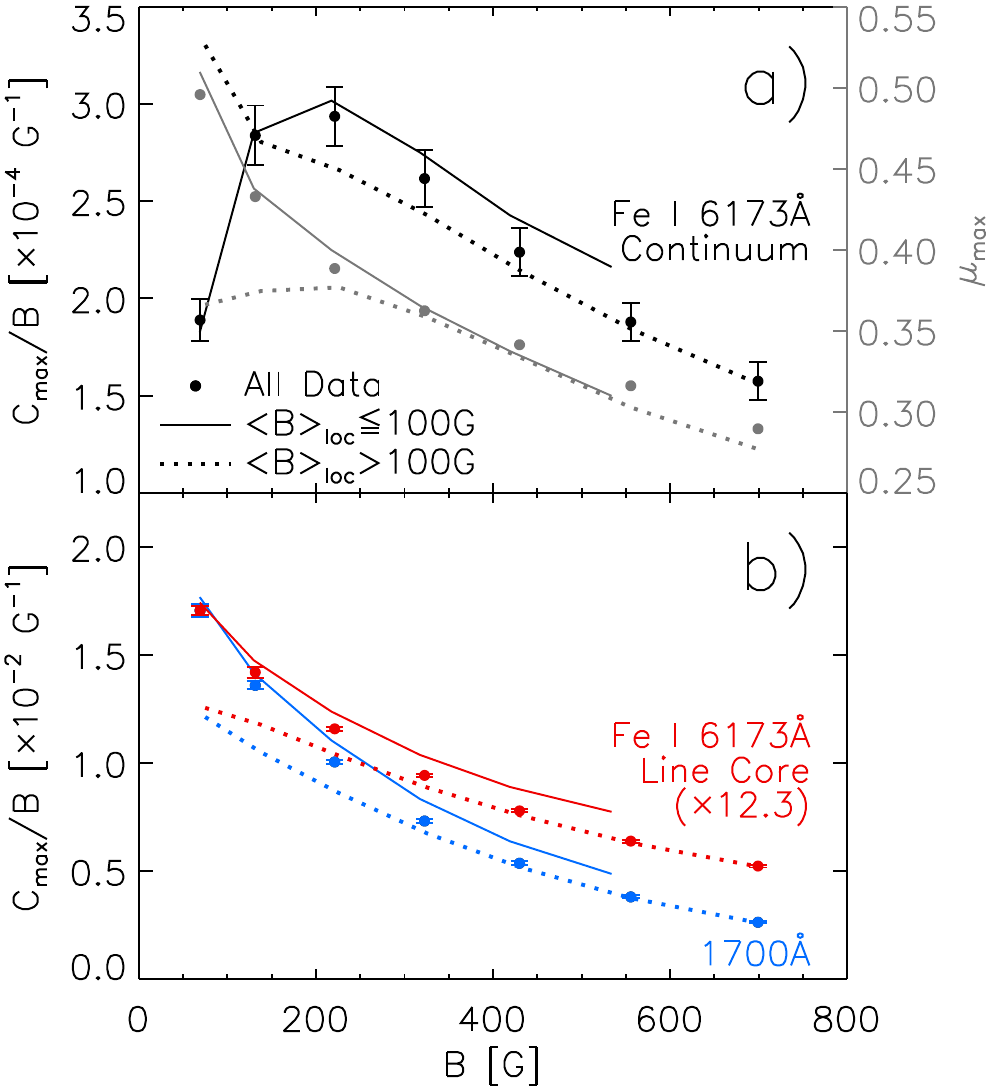}}
\caption{Same as Fig. \ref{contrastspecific}, except at $\bloc\leq100\ {\rm G}$ (solid curves) and $\bloc>100\ {\rm G}$ (dotted curves). The filled circles denote the values from taking all available data into consideration, adopted from the earlier figure. The uncertainty of the $\bloc<100\ {\rm G}$ and $\bloc>100\ {\rm G}$ series is of similar order as the ensemble series. It is omitted here to aid visibility.}
\label{contrastnbmuspecific}
\end{figure}

In this section, we examine the effect on the intensity contrast of the network and faculae through nearby magnetic fields, which are represented here by the local magnetic flux density, $\bloc$ (defined in Sect. \ref{analysis_loc}). To this end, we re-examine the CLV {at low and high $B$} (Fig. \ref{contrastnbmu}a), the $B$ dependence {near disc centre and the limb} (Fig. \ref{contrastnbmu}b), and the specific contrast (Fig. \ref{contrastnbmuspecific}) at the points where $\bloc\leq100\ {\rm G}$ (solid curves) and $\bloc>100\ {\rm G}$ (dotted curves) separately. We then compare the results to what we derived earlier in Sects. \ref{results_pix_a1} and \ref{results_pix_a2_pi}, for which we take all available data together (filled circles). At lower $B$, the values from the ensemble lie close to the $\bloc<100\ {\rm G}$ values and at higher $B$ to the $\bloc>100\ {\rm G}$ values. This reflects the fact that most of the points with lower (higher) $B$ are in the less (more) magnetically crowded network (faculae) fields.

First, we consider the continuum of the Fe I 6173 \AA{} line (black, Figs. \ref{contrastnbmu} and \ref{contrastnbmuspecific}). At low $B$ ($50\ {\rm G}<B\leq100\ {\rm G}$, Fig. \ref{contrastnbmu}a), the intensity contrast increases from $\bloc<100\ {\rm G}$ to $\bloc>100\ {\rm G}$ and peaks closer to the limb. At high $B$ ($500\ {\rm G}<B\leq640\ {\rm G}$, Fig. \ref{contrastnbmu}a), the intensity contrast is suppressed. The CLV profile peak is at about the same position, but the decline limb-wards of the peak is weaker such that the intensity contrast is higher near the limb. The effect of nearby magnetic fields on the intensity contrast varies gradually with $B$ between what is noted here at the lower and upper bounds, not shown for brevity. When we consider that the CLV is from the heated side walls of magnetic flux tubes coming into greater view before being blocked by the side facing away from the observer (cf. Sect. \ref{results_pix_a1_int}), the limb-ward shift of the CLV profile peak at low $B$ and the intensity contrast enhancement near the limb at high $B$ indicate greater flux tube sizes in magnetically crowded regions. 

Most of the studies of the continuum intensity contrast of the network and faculae at disc centre found that it declines into the negative as $B$ approaches 0 G \citep[see][and the summary of earlier results by Yea13]{kahil16}. Both here (Fig. \ref{contrastspecific}) and in Yea13, we find that below a certain level of $B$, the specific contrast in the continuum starts to decline. Both observations have been attributed to the tendency for magnetic concentrations to aggregate in the dark intergranular lanes (see the detailed argument in Yea13). Fig. \ref{contrastnbmu}b shows that as $B$ approaches 0 G, the intensity contrast is enhanced by the nearby magnetic fields. Below 150 G, the specific contrast at $\bloc<100\ {\rm G}$ (solid curve, Fig. \ref{contrastnbmuspecific}a) declines, as seen in the ensemble (filled circles), but the specific contrast at $\bloc>100\ {\rm G}$ (dotted curve) rises instead. We surmise that in magnetically crowded regions, the magnetic concentrations are overall less deeply embedded in intergranular lanes than in quieter regions.

Next we examine the Fe I 6173 \AA{} line core (red, Figs. \ref{contrastnbmu} and \ref{contrastnbmuspecific}) and 1700 \AA{} (blue).
At all levels of $B$, the intensity contrast from disc centre to limb changes from being suppressed to being enhanced by the local magnetic flux (Fig. \ref{contrastnbmu}b), such that the CLV is weaker (Fig. \ref{contrastnbmu}a). As noted in Sect. \ref{results_pix_a1_lcraia1700}, the decline in intensity contrast from disc centre to  the limb is in part due to intensity excess produced in magnetic flux tubes {becomes} absorbed into the interleaving non-magnetised atmosphere between flux tubes. For a given magnetic filling factor, the larger the flux tubes, the weaker this effect \citep{solanki98}. The weaker disc centre-to-limb drop is therefore another indication of greater flux tube sizes in magnetically crowded regions. In combination with the lower heating efficiency (Fig. \ref{contrastnbmuspecific}b), the intensity contrast switches from being suppressed to being enhanced by the local magnetic flux from disc centre to limb.

At each passband, except below about 100 G in the continuum of the Fe I 6173 \AA{} line, the specific contrast is higher at $\bloc<100\ {\rm G}$ (solid curves, Fig. \ref{contrastnbmuspecific}) than at $\bloc>100\ {\rm G}$ (dotted curves). As noted at the beginning of this section, from low to high $B$, the values calculated with the ensemble (filled circles) migrate from the level at $\bloc<100\ {\rm G}$ to the level at $\bloc>100\ {\rm G}$ because most of the lower (higher) $B$ points came from magnetically sparsely (crowded) regions. As asserted in Sect. \ref{results_pix_a2_pi}, the downward trend in heating efficiency with $B$ seen in the ensemble is partly due to the growing prevalence and influence of nearby magnetic fields.

{The analysis here revealed that even though the heating efficiency is largely suppressed by nearby magnetic fields, intensity contrast can be enhanced instead at certain $B$ and $\mu$. This is due to the effect of how magnetically crowded a given region is on flux tube sizes and how deeply embedded magnetic concentrations are in intergranular lanes.}

\subsection{Comparison of the quiet-Sun network and active region faculae}
\label{results_pix_b_lea93kea11}

\begin{figure*}
\sidecaption
\includegraphics[width=12cm]{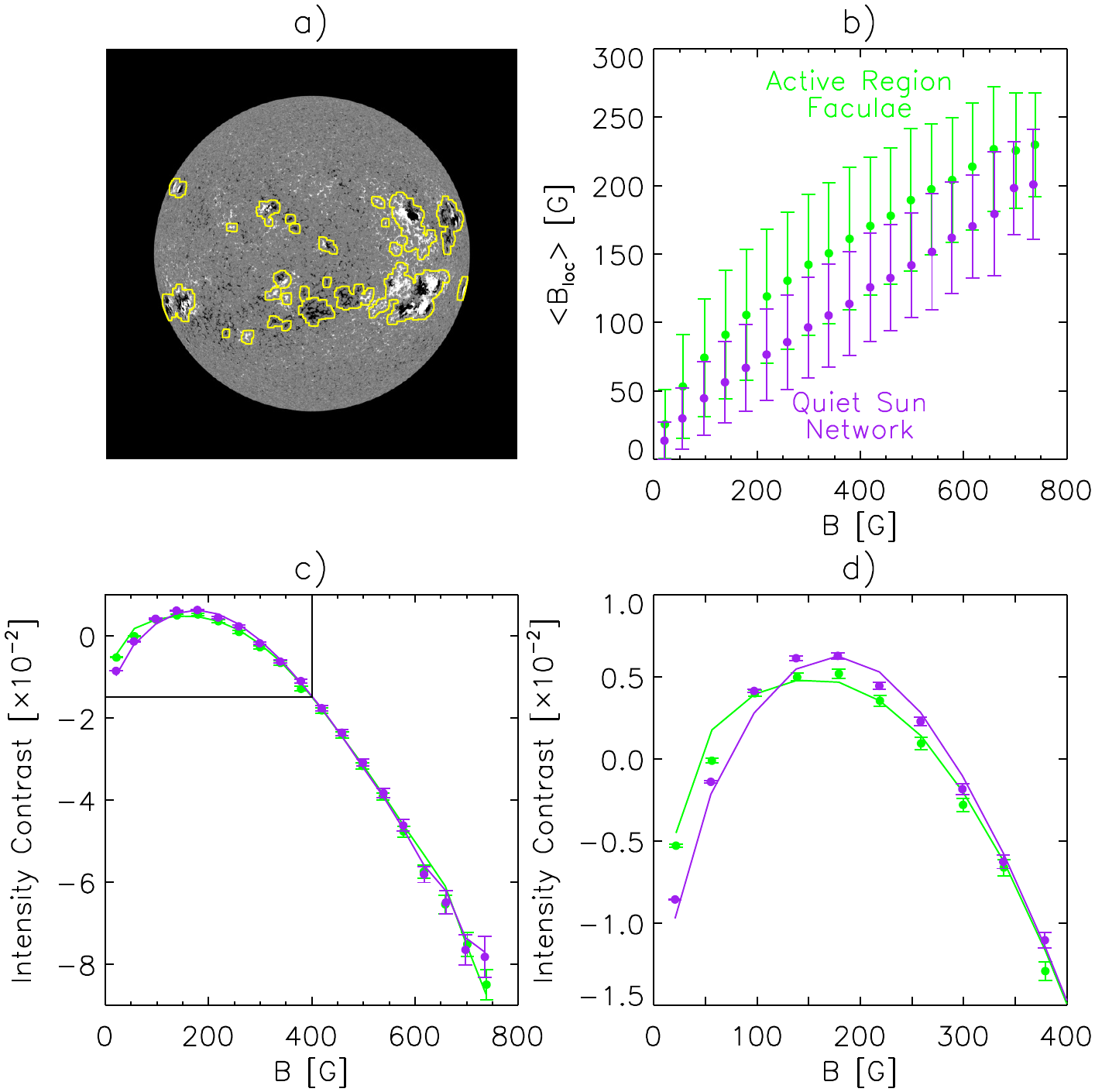}
\caption{a) Taken from Fig. \ref{detection}, the HMI longitudinal magnetogram from 12 June 2014. The yellow contours encompass the areas we identified as active regions. For the entire data set (i.e. all data days), {b)} $\bloc$ and {c)} the Fe I 6173 \AA{} continuum intensity contrast of the quiet-Sun network (purple) and active region faculae (green) at disc centre ($0.99<\mu\leq1.00$). {The filled circles represent the mean within successive intervals of 40 G. In b) the error bars denote the corresponding standard deviation, and in c) they denote the standard error.} The curves correspond to the contrast vs. $B$ profiles calculated with the empirical model described in Sect. \ref{results_pix_b_lea93kea11}. d) Zoomed inset of the boxed region.}
\label{contrastnbmudc}
\end{figure*}

\begin{figure*}
\sidecaption
\includegraphics[width=12cm]{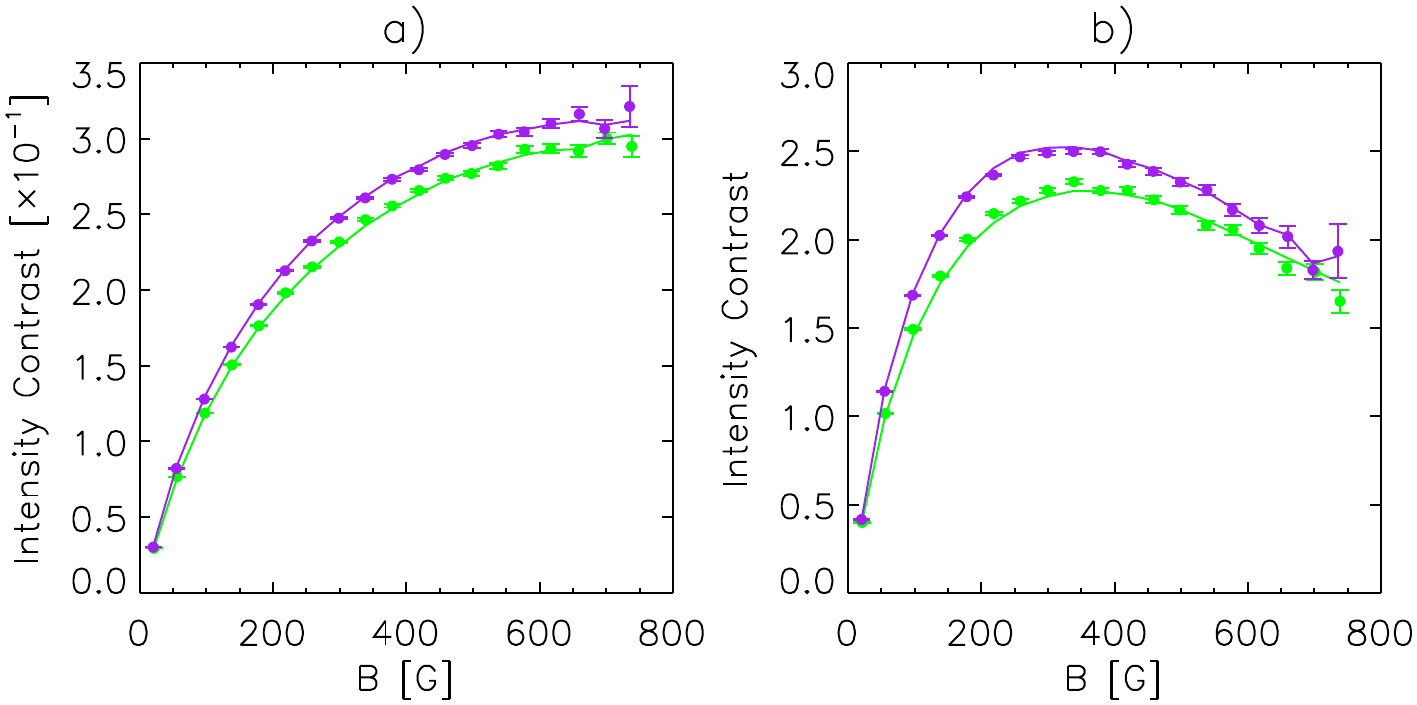}
\caption{Similar to Fig. \ref{contrastnbmudc}c, except for a) the Fe I 6173 \AA{} line core and b) 1700 \AA{}. {As in the earlier figure, the filled circles and error bars represent the mean and standard error within successive intervals of 40 G.}}
\label{contrastnbmudc2}
\end{figure*}

\begin{figure*}
\centering
\includegraphics[width=17cm]{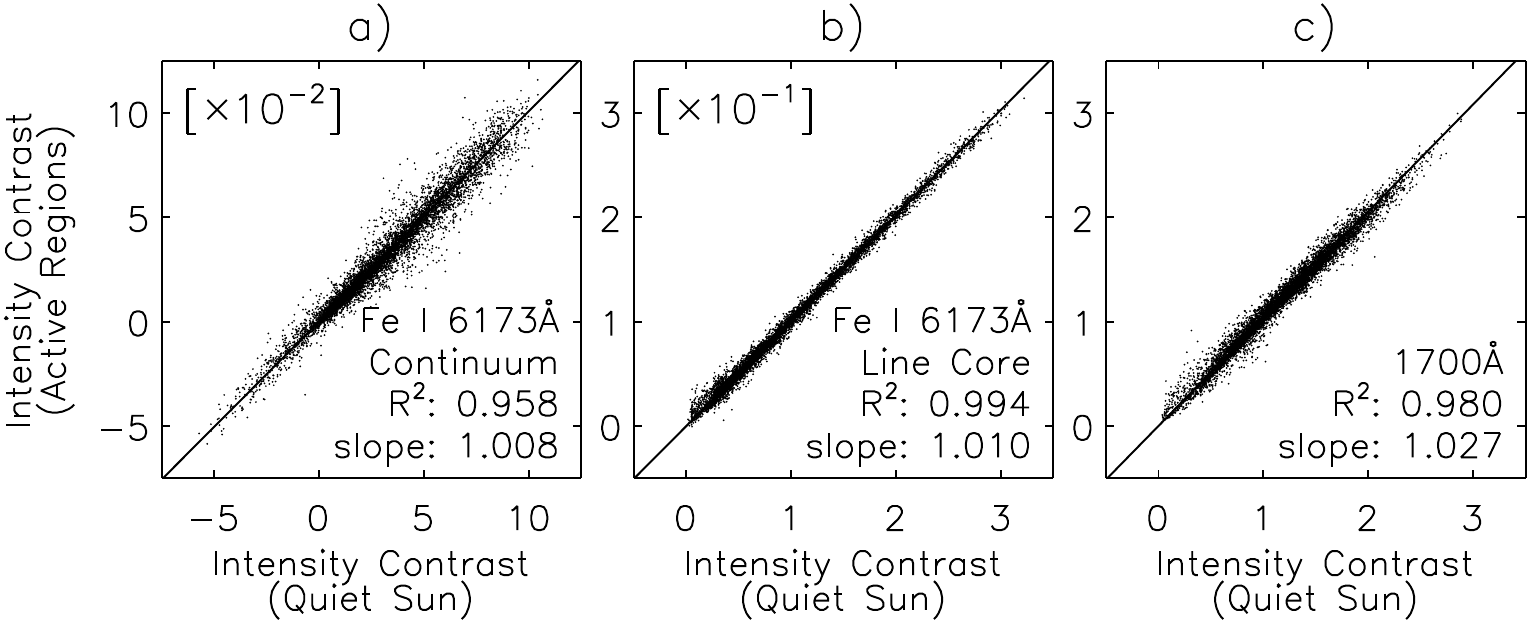}
\caption{Scatter plot of active region faculae vs. quiet-Sun network intensity contrast at a) the continuum and b) the core of the Fe I 6173 \AA{} line, and c) at 1700 \AA{}. The plot points represent the result of binning the intensity contrast measurements by $B$, $\bloc$ , and $\mu$, and taking the average within each bin (see Sect. \ref{results_pix_b_lea93kea11}). The straight lines denote the linear fit to each scatter plot, constrained to pass through the origin. The Pearson correlation coefficient $R^2$ and the slope of the linear fit are annotated.}
\label{contrastnbmuarqs}
\end{figure*}

Next, we investigate if the differences in intensity contrast between the quiet-Sun network and active region faculae noted by \cite{lawrence93} and \cite{kobel11}, discussed in the introduction, can be seen in the present data set, and if it can be accounted for by the effect of nearby magnetic fields on the intensity contrast alone. The implication being that if this were not the case, it implies that there is a fundamental physical difference between the quiet-Sun network and active region faculae.

For each data day, we segmented the solar disc into the quiet Sun and active regions, broadly following \cite{hagenaar03}. We refer to any contiguous patch of magnetically active pixels (i.e. where $\blos>3\sigma_{\blos}$) as a magnetic island. We isolated the magnetic islands where $\blos>5\sigma_{\blos}$ over an area of at least $11\times11$ pixels ($6.6\times6.6$ arcsec). Taking these magnetic islands to be the core of active regions, we expanded their boundary with a $51\times51$ pixel ($30.6\times30.6$ arcsec) boxcar filter. Taking the points isolated in Sect. \ref{analysis_fac}, we counted the points within the boundary as active region faculae and {outside the boundary} as quiet-Sun network. An example of the segmentation is depicted in Fig. \ref{contrastnbmudc}a. As expected, the distribution of $\bloc$ is higher in active region faculae than in the quiet-Sun network (Fig. \ref{contrastnbmudc}b).

Following \cite{kobel11}, we compared the continuum intensity contrast in the quiet Sun (purple, Figs. \ref{contrastnbmudc}c and \ref{contrastnbmudc}d) and in active regions (green) at disc centre ($0.99<\mu\leq1.00$). The result is qualitatively similar to what was reported by the earlier study; compared to active region faculae, the quiet-Sun network intensity contrast is higher about the peak of the contrast versus $B$ profile, and lower approaching $B=0\ {\rm G}$. Taking all the points identified as network and faculae at $0.99<\mu\leq1.00$, we binned the points by $B$ and $\bloc$ (in intervals of 20 G) and fit a quartic bivariate polynomial in $B$ and $\bloc$ to the bin-averaged intensity contrast. The fit is in effect an empirical model of disc-centre intensity contrast as a function of $B$ and $\bloc$ alone (i.e. assuming no fundamental physical difference between network and faculae). We re-derived the contrast versus $B$ profiles by calculating the intensity contrast of each network and facular image pixel from its $B$ and $\bloc$ with this empirical model instead of taking the apparent contrast in the intensity image. As evident in Figs. \ref{contrastnbmudc}c and \ref{contrastnbmudc}d, the modelled profiles (curves) are in close agreement with the measured profiles (filled circles). In Fig. \ref{contrastnbmudc2} we show similar results from repeating this analysis on the Fe I 6173 \AA{} line core and 1700 \AA{}. The results here suggest that the apparent divergence between the quiet-Sun network and active region faculae intensity contrast arises is because active regions are more magnetically crowded (Fig. \ref{contrastnbmudc}b). {If the apparent divergence between quiet Sun and active region intensity contrast were statistical or would arise from factors other than the local magnetic flux density, then the modelled profiles should overlap with one another instead of reproducing the difference between the measured quiet-Sun and active region profiles.} This is supported by the following.

We binned the points by $B$, $\bloc$ , and $\mu$ (in intervals of 20 G, 20 G, and 0.25, respectively) separately for the quiet-Sun network and active region faculae, and took the bin-averaged intensity contrast. The scatter plot of active region faculae versus quiet-Sun network bin-averaged intensity contrast at each passband is depicted in Fig. \ref{contrastnbmuarqs}. The correlation (Pearson's $R^2$) between network and facular intensity contrast and the slope of the straight-line fit to the scatter plot, indicated in the figure, are within about 1\% to 4\% of unity. In other words, after factoring out differences in $B$, $\bloc$ , and $\mu$, the network and facular intensity contrast are similar to this margin.

There is no indication, at least within the limits of this study, of any fundamental physical difference between network and faculae. The different radiant behaviour of the quiet-Sun network and active region faculae arise because the faculae face more nearby magnetic fields alone. The results here and in Sect. \ref{results_pix_b} highlight the need for models of solar irradiance to include the effect of nearby magnetic fields on the network and facular intensity. This is so far lacking in existing models.

\subsection{Comparison of the Fe I 6173 \AA{} line core and 1700 \AA{}}
\label{results_pix_c}

\begin{figure*}
\centering
\includegraphics[width=17cm]{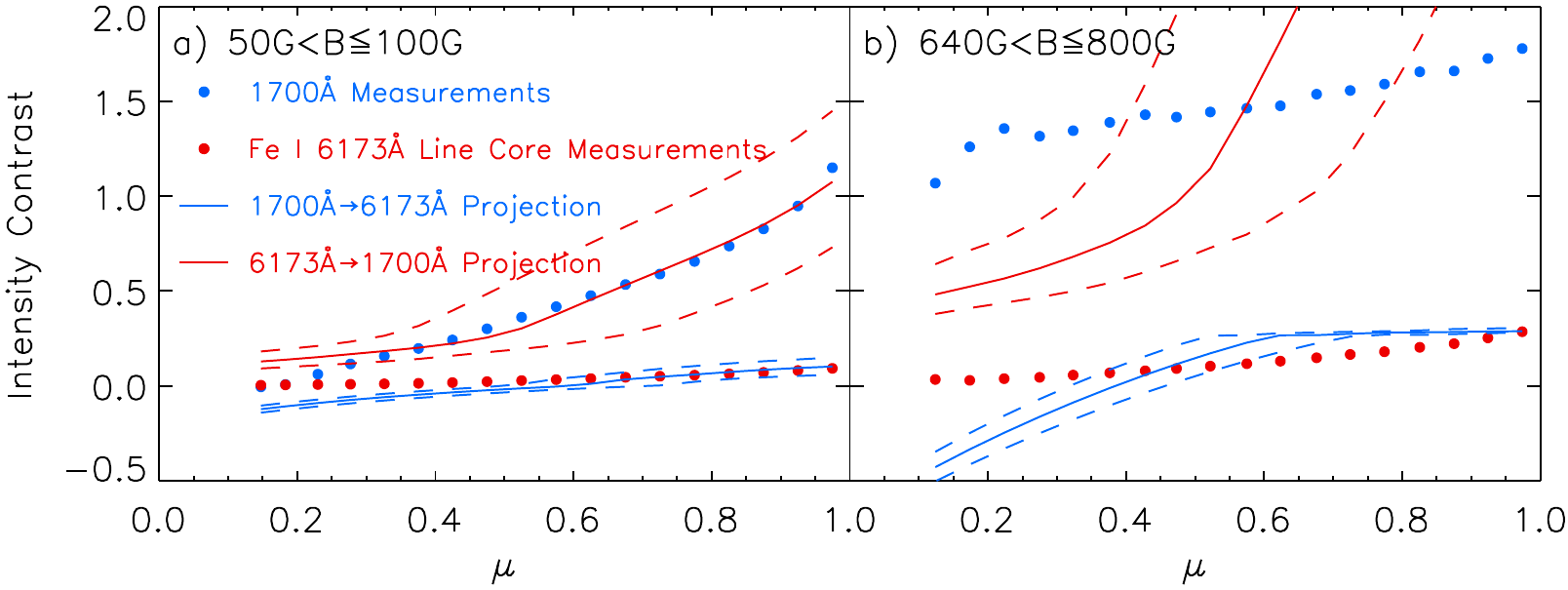}
\caption{Filled circles: Intensity contrast CLV at the Fe I 6173 \AA{} line core (red) and 1700 \AA{} (blue), at a) $50\ {\rm G}<B\leq100\ {\rm G}$ and b) $640\ {\rm G}<B\leq800\ {\rm G}$, taken from Fig. \ref{contrastbinbybmu}b. Curves: Result of estimating the intensity contrast at one passband from the measurements at the other, assuming the solar atmosphere radiates like a black body {(solid) and the corresponding 1$\sigma$ bounds from the uncertainty in the effective temperature at 1700 \AA{} (dashed).} See Sect. \ref{results_pix_c}.}
\label{contrastprojection}
\end{figure*}

Here, we investigate what drives the divergence between intensity contrast at the Fe I 6173 \AA{} line core and 1700 \AA{} in addition to the Zeeman line changes in the former (discussed in Sects. \ref{results_pix_a1} and \ref{results_pix_a2_pi}).
We study the intensity contrast CLV at $50\ {\rm G}<B\leq100\ {\rm G}$ (Fig. \ref{contrastprojection}a). Assuming the solar atmosphere radiates like a black body, we estimated from the measured intensity contrast at 1700 \AA{} ({blue circles}) the intensity contrast at the Fe I 6173 \AA{} line core ({blue curve}). We also projected the measured intensity contrast at the Fe I 6173 \AA{} line core ({red circles}) to 1700 \AA{} ({red curve}). This calculation requires as input the effective temperature of the quiet Sun at both passbands. We assumed an effective temperature of 5800 K at the Fe I 6173 \AA{} line core. We then set the effective temperature at 1700 \AA{} at 6020 K; this is the level that optimises the agreement between measured and projected intensity contrast. (The effective temperature of the quiet Sun at the two passbands cannot be identical given the difference in formation height, cf. Sect. \ref{data}.) This analysis, however approximate, shows an excellent agreement between measured and projected intensity contrast.

We confine this discussion to the low $B$ regime because this is where the Zeeman line changes are weakest and can be factored out. For information, we depict the results of repeating the above calculation at high $B$ in Fig. \ref{contrastbinbybmu}b. As expected, the projection, which ignores the Zeeman line changes, fails to reproduce the measured intensity contrast. In addition to the Zeeman line changes, the divergence between the intensity contrast at the Fe I 6173 \AA{} line core and 1700 \AA{} arises from the different wavelength and temperature regime (from the different formation height), as governed by the black-body law.

Photospheric iron lines, including the Fe I 6173 \AA{}, are formed largely in LTE. Departures from LTE are sufficiently minute such that observed line profiles can be equally well replicated by LTE and non-LTE radiative transfer schemes with an adjustment of the iron abundance \citep{rutten82,bruls91,shchukina01}. On the other hand, it is not understood whether the AIA 1700 \AA{} channel is formed in LTE or non-LTE, due in part to the forbiddingly complicated zoo of relevant species \citep[see, e.g.][]{fossum05}. Therefore, the fact that we can replicate intensity contrast at one passband with the measurements from the other assuming black-body behaviour {suggests} either that the AIA 1700 \AA{} channel is formed in LTE or that the AIA 1700 \AA{} channel is formed in non-LTE, but departures from LTE affect the non-magnetised and magnetised solar atmosphere in such a way that the intensity contrast is  not affected. {As the effective temperature at 1700 \AA{} is a fit parameter, set at the level that optimises the agreement between measured and modelled intensity contrast, the agreement between the two could also mean that the AIA 1700 \AA{} channel is formed in non-LTE and non-LTE effects on the intensity contrast can be reproduced by assuming the appropriate effective temperature.}

{If indeed departures from LTE affect the non-magnetised and magnetised solar atmosphere in such a way that the intensity contrast is unaffected,} it will explain the following: Certain models of solar irradiance \citep{krivova06,ball11,yeo14b} make use of the intensity contrast of solar surface magnetic features that are calculated from model solar atmospheres assuming LTE \citep{unruh99}. The \cite{yeo14b} model reproduces the observed variability in the wavelength-resolved or spectral solar irradiance all the way down to about 160 nm even though the absolute level already {starts to differ from} measurements at below 300 nm {as} the LTE assumption breaks down (\cite{yeo14b}.

\subsection{Comparison with the Ca II K line}
\label{results_pix_d}

\begin{figure*}
\centering
\includegraphics[width=17cm]{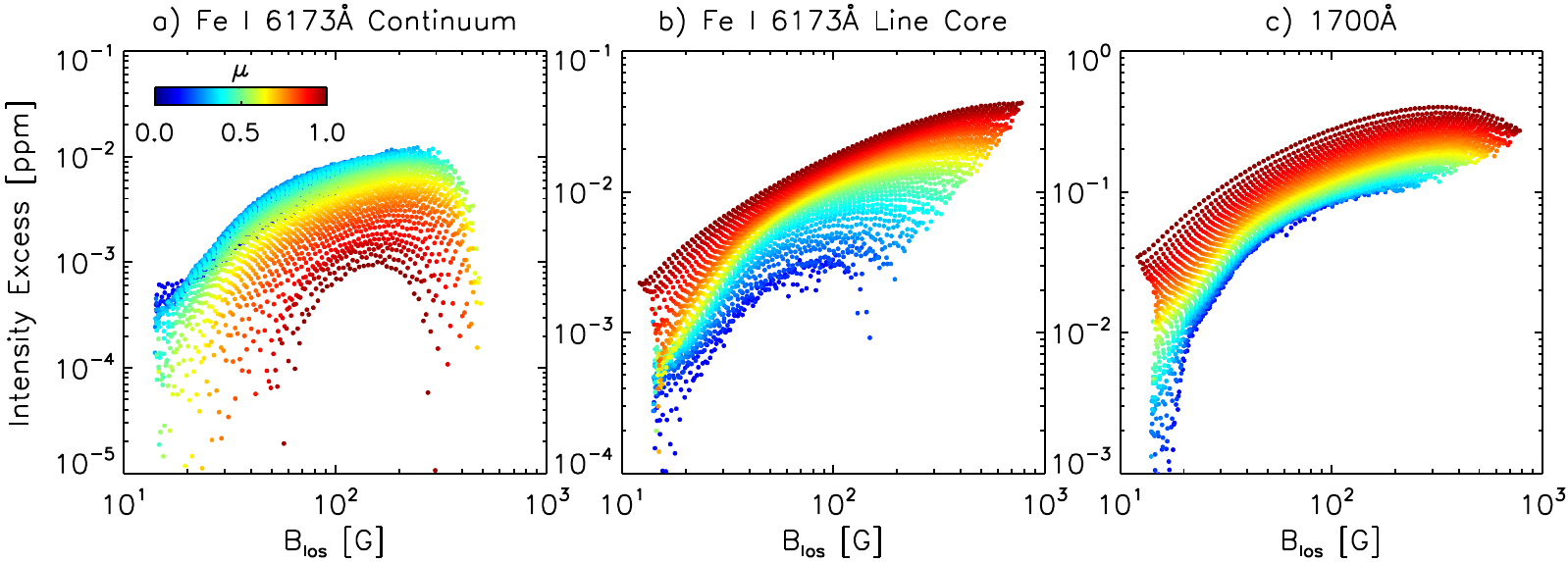}
\caption{Log-log scatter plot of the intensity excess and $\blos$ at a) the continuum and b) the core of the Fe I 6173 \AA{} line, and c) 1700 \AA{}. The intensity excess is given in units of ppm of the disc-integrated quiet-Sun intensity. The plot points, colour-coded by $\mu$, denote the result of grouping the network and facular image pixels with similar $\blos$ and $\mu$ together, and taking the average intensity excess. See Sect. \ref{results_pix_d}.}
\label{scalingmagflux}
\end{figure*}

\begin{figure*}
\centering
\includegraphics[width=17cm]{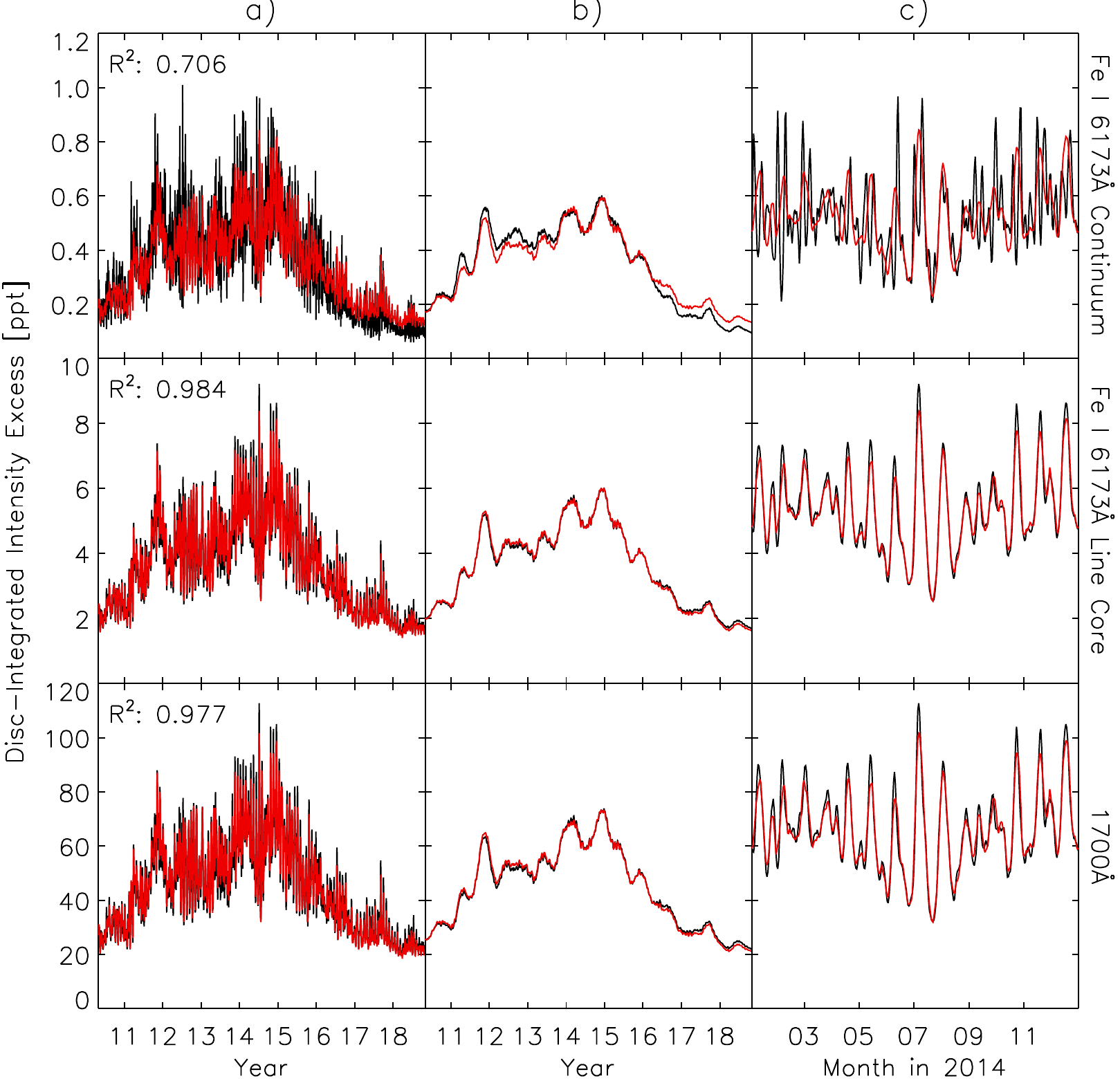}
\caption{a) Black: Daily disc-integrated intensity excess at the continuum and core of the Fe I 6173 \AA{} line and 1700 \AA\ (top to bottom){}, calculated with the empirical intensity contrast model derived in this study (Equation \ref{surfacefiteqn} and Table \ref{surfacefit}). Red: Fit of the daily disc-integrated intensity excess at the Ca II K line, calculated with the relationship between Ca II K intensity excess and $\blos$ reported by \cite{harvey99}. The $R^2$ between the two time series is annotated. b) The 81-day moving average. c) Zoomed inset of 2014.}
\label{contrastchromo}
\end{figure*}

Various studies have reported power-law relationships between chromospheric emission and magnetic flux density \citep{schrijver89,harvey99,ortiz05,rezaei07,loukitcheva09,kahil16}. Notably, the study by \cite{harvey99}, the only to cover the full solar disc, found the relationship between intensity excess at the Ca II K line, the difference in intensity to the quiet-Sun level, and $\blos$ to be invariant with $\mu$. We investigate if emission at the passbands examined in this study, formed in the lower and middle photosphere (Sect. \ref{data}), conform to similar relationships with $\blos$ and $\mu$. In keeping with the cited studies, {we study the relationship between intensity excess, denoted $\Delta{}I$, and $\blos$, instead of intensity contrast and $B,$ as in the rest of this study.}

In Fig. \ref{scalingmagflux} we show the log-log scatter plot of $\Delta{}I$ and $\blos$ at each passband. A power-law relationship between $\Delta{}I$ and $\blos$ that is invariant with $\mu$ will show itself in such a representation as the plot points, coming from across the entire disc, falling along the same straight line. As is evident from the figure, $\Delta{}I$ does not conform to a power-law relationship with $\blos$, and neither is the relationship with $\blos$ invariant with $\mu$ at any of the passbands.

What is the implication for models of solar irradiance that use chromospheric indices, including indices based on the Ca II K line \citep{keil98}, as direct indications of the effect of network and facular on solar irradiance? The use of chromospheric indices as proxies of network and facular brightening is based on the fact that they are dominantly modulated by the chromospheric network and plage overlaying photospheric network and faculae. With this in consideration, we examine the variation in disc-integrated intensity, from the changing prevalence and distribution of network and faculae, at the passbands examined here and at the Ca II K line. 

The HMI has been in operation since 30 April 2010. For each day since, up to 31 October 2018, we took a 315s longitudinal magnetogram and continuum intensity image, and identified the network and faculae as done in Fig. \ref{analysis_fac}, except that we did not exclude the non-sunspot magnetic activity contiguous to sunspots and pores (this makes no difference to the current analysis). At each passband, using the empirical model of intensity contrast as a function of $B$ and $\mu$ derived in Sect. \ref{results_pix_a1_er}, we determined the daily disc-integrated network and facular intensity excess (black, Fig. \ref{contrastchromo}a). To elucidate the variability over solar cycle and rotation timescales, we also draw the 81-day moving average (Fig. \ref{contrastchromo}b) and the zoomed inset of 2014 (Fig. \ref{contrastchromo}c).

\cite{harvey99} reported a power-law exponent of 0.5, or in other words, $\Delta{}I\propto\sqrt{\blos}$ at the Ca II K line. For each day, we took the sum of $\sqrt{\blos}$ of the network and facular image pixels. The result is a time series with the variability, although not the absolute scale, of the disc-integrated network and facular intensity excess at the Ca II K line. We fit this time series to each of the three other disc-integrated intensity excess time series (red, Fig. \ref{contrastchromo}). The studies similar to \cite{harvey99}, cited in the beginning of this section, reported power-law exponents ranging from about 0.1 to 0.7. Tests showed that varying the power-law exponent adopted in the this calculation within this range {makes no appreciable difference} the result.

The disc-integrated intensity excess fluctuates as network and faculae rotate across the solar disc due to the CLV of the intensity contrast/excess. At the various passbands examined in this study, the CLV of the intensity excess is qualitatively similar to the CLV of the intensity contrast (Fig. \ref{contrastbinbybmu}), which is not drawn here to avoid repetition. Like the intensity contrast, the intensity excess at the Fe I 6173 \AA{} continuum exhibits the opposite CLV as the Fe I 6173 \AA{} line core and 1700 \AA{}, giving the discrepant rotational variability in the disc-integrated intensity excess (Fig. \ref{contrastchromo}c).

The Fe I 6173 line core and 1700 \AA{} time series are in close agreement with the Ca II K line time series, which is evident in the alignment over solar cycle (Fig. \ref{contrastchromo}b) and rotation timescales (Fig. \ref{contrastchromo}c), and in the $R^2$ (0.984 and 0.977, respectively). This even though the Fe I 6173 \AA{} line core and 1700 \AA{} emission do not conform to a similar relationship with $\blos$ and $\mu$ as Ca II K emission (Fig. \ref{scalingmagflux}). The converse is true at the Fe I 6173 \AA{} continuum, especially at solar rotation timescales (Fig. \ref{contrastchromo}c).

From $\Delta{}I\propto\sqrt{\blos}$ at the Ca II K line and $B=\blos/\mu$, it is straightforward to show that at a given level of $B$, the Ca II K intensity excess scales with $\sqrt{\mu}$. Similar to the Fe I 6173 \AA{} line core and 1700 \AA{}, and opposite to the Fe I 6173 \AA{} continuum, the Ca II K intensity excess declines from disc centre to the limb. In terms of the time variability, the Ca II K time series is therefore much closer to the Fe I 6173 \AA{} line core and 1700 \AA{} time series than the Fe I 6173 \AA{} continuum time series. We recall that the intensity contrast/excess CLV of the Fe I 6173 \AA{} continuum is opposite to that of the Fe I 6173 \AA{} line core and 1700 \AA{} because of the different physical processes that underlie the magnetic enhancement of the intensity in the lower and middle photosphere (Sects. \ref{results_pix_a1_int} and \ref{results_pix_a1_lcraia1700}). The results here indicate that even though the lower and middle photospheric emission does not scale with magnetic flux density and distance from disc centre in the same manner as the chromospheric emission (Fig. \ref{scalingmagflux}), the time variation in the chromospheric emission from the network and faculae remains reasonably similar to that in the middle photospheric emission but not to that in the lower photosphere emission (Fig. \ref{contrastchromo}) because of the way in which the CLV of the intensity contrast/excess in the various atmospheric regimes compare. This represents a fundamental limit to the usefulness of chromospheric indices as direct indications of the effect of network and faculae on solar irradiance.

{The disagreement between the Ca II K and Fe I 6173 \AA{} continuum time series is caused by the opposite CLV of the intensity contrast/excess at the two passbands. In this calculation, we assumed the overall relationship between Ca II K intensity excess and $\blos$ reported by \cite{harvey99}. The authors also noted that when they considered the various solar surface magnetic feature types separately (such as enhanced network and active region plage), the power-law coefficients, that is, the constant of proportionality and the exponent, differ between the various feature types. The power-law exponent that \cite{harvey99} reported for the various feature types they examined ranges from 0.4 to 0.8. This means that at a given level of $B$, the Ca II K intensity excess scales with $\mu^{0.4}$ to $\mu^{0.8}$ depending on feature type, declining from disc centre to limb as in the overall case. We have neglected the variation in the power-law coefficients with feature type when we calculated the Ca II K time series. This likely has only very little qualitative {effect} on how the result compares to the other time series.}

\subsection{Variation with time}
\label{results_pix_e}

\begin{figure*}
\centering
\includegraphics[width=17cm]{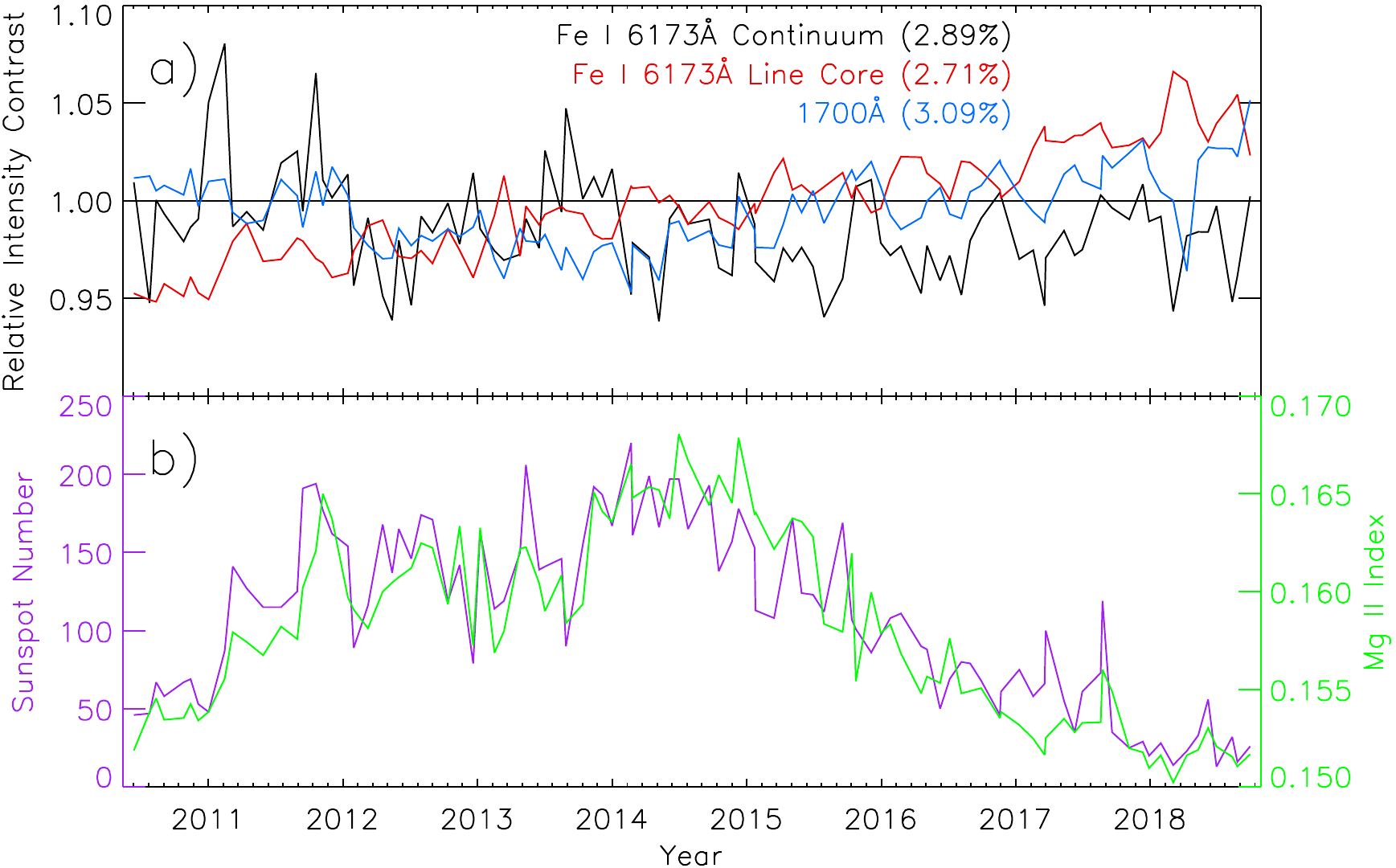}
\caption{a) Scale of the intensity contrast on each data day relative to the mean level over all the data days at the core (black) and continuum (red) of the Fe I 6173 \AA{} line, and 1700 \AA{} (blue). The RMS deviation from unity is indicated. b) Solar activity at the data days, as indicated by the sunspot number (purple) and the Mg II index (green).}
\label{contraststability}
\end{figure*}

The 100 data days cover a nine-year period that encompasses the maximum of solar cycle 24,\footnote{The maximum of solar cycle 24, as indicated by the 13-month running mean of the monthly sunspot number, was reached in April 2014.} giving us the opportunity to investigate if the intensity contrast of network and faculae varies with cycle phase.
For each passband and each data day, we binned the intensity contrast measurements by $\mu$, $B,$ and $\bloc$ (in intervals of 0.25, 20 G, and 20 G, respectively) and took the binned average. We fit a straight line to the scatter plot of the binned average intensity contrast and the mean over all data days. The slope of the straight line fit gives the scale of the intensity contrast on the data day relative to the mean over all data days, expressed in Figs. \ref{contraststability}a. We refer to this quantity as the relative intensity contrast. The relative intensity contrast time series reveals that the intensity contrast is stable with time to about $3\%$. More importantly, there is no common trend between the various time series, nor do they indicate any clear correlation to solar cycle phase (by the comparison to the sunspot number and the Mg II index, Fig. \ref{contraststability}b). Within the limits of our study, the intensity contrast of network and faculae does not appear to vary with cycle phase.

\cite{ortiz06} repeated the analysis of \cite{ortiz02} on MDI data covering 1996 to 2001, a six-year period spanning the rising phase of solar cycle 23. They studied the intensity contrast of network and faculae on a year-to-year basis and found it to be stable to about 10\% over this period. Not only is this margin of $10\%$ non-trivial, it was not known at the time that the instrumental response of MDI might have changed over the extended outages that occurred between June 1998 and February 1999 \citep{ball12}. The finding here that HMI and AIA, two independent instruments, indicate that the intensity contrast of network and faculae is stable to a much smaller margin of $3\%$ over a longer period that encompasses almost an entire solar cycle represents a significant advance over this earlier work.

\section{Summary}
\label{summary}

We examined the intensity contrast of network and faculae in the continuum and core of the Fe I 6173 \AA{} line and the AIA 1700 \AA{} channel. We employed simultaneous full-disc observations of the magnetic flux density and the intensity at the various passbands from the HMI and AIA instruments on board SDO. The data set comprises observations from 100 days spanning June 2010 to October 2018. We examined the variation in intensity contrast with distance from disc centre and magnetic flux density, $B$, the average magnetic flux density in the greater area about each point, called the local magnetic flux density, $\bloc$, and time.

This is the first study of its kind to examine the variation in intensity contrast with local magnetic flux density over the entire solar disc. This was prompted by the observation that the quiet-Sun network appears for the same amount of magnetic flux to be brighter than active region faculae, a phenomena attributed to the {fact} that active regions are more magnetically crowded. We note that even though the heating efficiency of the magnetic flux is within expectation suppressed by nearby magnetic fields as a result of the effect of how magnetically crowded a given region is on flux tube sizes and how deeply embedded the magnetic concentrations are in intergranular lanes, the intensity contrast is  enhanced at certain levels of $B$ and distances from disc centre. Using an empirical model of the intensity contrast as a function of distance from disc centre, $B$ and $\bloc$, derived here from the intensity contrast measurements, we reproduced the difference in intensity contrast between the quiet-Sun network and active region faculae. This means that the difference arises because active regions are more magnetically crowded alone and is not due to any fundamental physical difference between network and faculae. In addition, by comparing network and faculae at similar distances from disc centre, $B$ and $\bloc$, we found that the network and facular intensity contrast is similar to about $1\%$. This confirms that network and faculae are not fundamentally different from one another. The results here highlight that models of solar irradiance need to include the effect of nearby magnetic fields on the network and facular intensity. This is so far lacking in existing models.

It is understood that the Fe I 6173 \AA{} line is formed largely in LTE, but it is not certain if the AIA 1700 \AA{} channel is formed in LTE or non-LTE. We found that we can estimate the intensity contrast at one passband from the measurements at the other, assuming that the solar atmosphere radiates like a black body. This means either that the AIA 1700 \AA{} channel is also formed in LTE, or that it is formed in non-LTE and that departures from LTE have little effect on the intensity contrast. If the latter is true, it would explain why models of solar irradiance that are based on the intensity contrast of solar surface magnetic features that are calculated assuming LTE can reproduce the observed variability in spectral solar irradiance even where the LTE assumption breaks down.

Chromospheric indices are employed in certain models of solar irradiance as direct indications of the effect of network and faculae on solar irradiance. The physical basis is the fact that chromospheric indices are largely modulated by the chromospheric network and plages that overlie the photospheric network and faculae. To investigate the validity of this approach, we calculated the daily disc-integrated intensity excess from network and faculae over the period of April 2010 to October 2018 in the passbands examined here (using an empirical model based on the intensity contrast measurements) and in the Ca II K line \citep[adopting the relationship between Ca II K intensity and magnetic flux density reported by][]{harvey99}. In terms of variation with time, the Ca II K time series is in close agreement with the Fe I 6173 \AA{} line core and 1700 \AA{} time series, but not with the Fe I 6173 \AA{} continuum time series. The reason is that the CLV of the intensity excess at the Ca II K line is opposite to that at the Fe I 6173 \AA{} continuum, but is qualitatively similar to that at the Fe I 6173 \AA{} line core and 1700 \AA{}. This divergence stems from the different physical processes that underlie the magnetic enhancement of the intensity in the lower photosphere, where the Fe I 6173 \AA{} continuum is formed, and in the middle photosphere, where the Fe I 6173 \AA{} line core and the AIA 1700 \AA{} channel are formed. The results here indicate that chromospheric indices might be a reasonable proxy of the time variability in middle photospheric emission that is due to network and faculae, but not that in lower photosphere emission (Fig. 14). This represents a fundamental limit of solar irradiance models that employ chromospheric indices as a proxy of network and facular brightening.

By comparing the intensity contrast of the network and faculae on each data day to the average over all data days, we found that the intensity contrast is stable over the nine-year period extended by the data set to about $3\%$. This is significantly lower than the margin of $10\%$ reported by \cite{ortiz06}. More importantly, there is no indication of any relationship between intensity contrast and solar cycle phase. This is of direct relevance to models of solar irradiance that are based on the calculated intensity contrast of solar surface magnetic features, where it is assumed that the intensity contrast is invariant with time.
        
The results of this study offer new and valuable insights into the radiant behaviour of network and faculae. These are of direct relevance to solar irradiance modelling.

\begin{acknowledgements}
We made use of observations from the AIA and HMI instruments on board the SDO mission (available at jsoc.stanford.edu), the SILSO sunspot number record (www.sidc.be/silso/), and the IUP Mg II index composite (www.iup.uni-bremen.de/gome/gomemgii.html). This study was supported by the German Federal Ministry of Education and Research under project 01LG1209A.
\end{acknowledgements}

\bibliographystyle{aa}
\bibliography{references}

\end{document}